\documentclass[reprint, showpacs, nofootinbib, nobibnotes, amsmath,amssymb, aps, showkeys]{revtex4-1}

\usepackage{graphicx}
\usepackage{dcolumn}
\usepackage{bm}
\usepackage{subfigure}
\usepackage{soul}
\usepackage{color}

\begin{document}

\preprint{APS/123-QED}

\title{Stretching of viscoelastic drops by steady sliding}
\author{Silvia Varagnolo}
\altaffiliation{present address: PDRA Department of Chemistry - University of Warwick, United Kingdom CV4 7AL}
\affiliation{
Dipartimento di Fisica e Astronomia ``G. Galilei'' - DFA and Sezione CNISM,
Universit\`a di Padova,
Via Marzolo 8, 35131 Padova (PD), Italy
}

\author{Daniele Filippi}
\affiliation{
Dipartimento di Fisica e Astronomia ``G. Galilei'' - DFA and Sezione CNISM,
Universit\`a di Padova,
Via Marzolo 8, 35131 Padova (PD), Italy
}

\author{Giampaolo Mistura}
\affiliation{
Dipartimento di Fisica e Astronomia ``G. Galilei'' - DFA and Sezione CNISM,
Universit\`a di Padova,
Via Marzolo 8, 35131 Padova (PD), Italy
}

\author{Matteo Pierno}
\email{matteo.pierno@unipd.it}
\affiliation{
Dipartimento di Fisica e Astronomia ``G. Galilei'' - DFA and Sezione CNISM,
Universit\`a di Padova,
Via Marzolo 8, 35131 Padova (PD), Italy
}

\author{Mauro Sbragaglia}
\email{sbragaglia@roma2.infn.it}
\affiliation{Dipartimento di Fisica, Universit\`a di Roma  ``Tor Vergata'' and INFN , Via della Ricerca Scientifica, 1 - 00133  Roma  RM (Italy)
}

\date{\today}

\begin{abstract}
The sliding of non-Newtonian drops down planar surfaces results in a complex, entangled balance between interfacial forces and non linear viscous dissipation, which has been scarcely inspected. In particular, a detailed understanding of the role played by the polymer flexibility and the resulting elasticity of the polymer solution is still lacking.
To this aim, we have considered polyacrylamide (PAA) solutions of different molecular weights, suspended either in water or glycerol/water mixtures. 
In contrast to drops with stiff polymers,
drops with flexible polymers exhibit a remarkable elongation in steady sliding.
This difference is most likely attributed to different viscous bending as a consequence of different shear thinning.
Moreover, an ``optimal elasticity'' of the polymer seems to be required for this drop elongation to be visible. We have complemented experimental results with numerical simulations of a viscoelastic FENE-P drop. This has been a decisive step to unravel how a change of the elastic parameters (e.g. polymer relaxation time, maximum extensibility) affects the dimensionless sliding velocity.
\end{abstract}

			
\maketitle



\section{Introduction}\label{intro}

Controlling and manipulating drops on open surfaces is a crucial step for applications in many fields, including chemistry, biomedicine, ink-jet printing, food and pharmaceutical industry \cite{quere08,Bonn09,Liu10,Brunet16,Nakajima16,Semprebon16}. The vast majority of such applications involves non-Newtonian fluids, e.g. polymer solutions, biological samples \cite{wheeler10,mousa09}, blood \cite{laan14,wheeler12} or inks \cite{son09}, characterised by a non-linear response to external stresses, consisting either in a shear dependent viscosity (shear-thinning, shear-thickening or yield stress fluids) or in elastic effects related to the appearance of normal stresses. Despite the wide spread of complex fluids, the research about the behaviour of non-Newtonian drops is quite recent and limited mainly to spreading  \cite{Rafai05,Rafai,Carre97,HanKim13,min10}, dynamic wetting \cite{wei09,wei07,wei07thinning,wang07,boudaoud07,HanKim14,liu09,min13,Nakajima11}, and impacting \cite{boyer16,zang14,zang13,ravi13,an12,huh15} on surfaces. To the best of our knowledge, only a few works \cite{EPJnoi,Morita09} analyse the dynamics of non-Newtonian drops sliding down an inclined surface. Notably, polymer solutions usually display both shear-thinning and elasticity, which are somehow distinctive of two broad categories of non-Newtonian behaviours, in principle with different phenomenology. To decouple these non-Newtonian effects, existing studies \cite{wei07,wei07thinning,wei09} consider polymeric fluids mainly featuring either shear-thinning or elasticity. Spreading measurements \cite{Rafai05,Rafai,Carre97,HanKim13,min10} report very weak deviations from the Newtonian case, both for solutions having a shear-dependent viscosity and for normal stress fluids, suggesting that non-Newtonian features are not relevant in this kind of phenomena. On the other hand, dynamic wetting \cite{wei09} observed by moving a solid tube in a cylinder of liquid is characterised by strong non-Newtonian deviations for shear-thinning solutions and weaker effects in case of normal stress fluids. Shear-thinning induces a reduction of viscous bending near the contact line and a lower drag of the solid surface on the fluid in the wedge like region with respect to a Newtonian reference having the same zero-shear viscosity \cite{wei07,Seevaratnam07}. Elasticity present in Boger fluids promotes an increase of curvature near the contact line, even if differences between Newtonian and non-Newtonian liquids are very weak \cite{wei09}. However the strongest differences between Newtonian and non-Newtonian drops are observed in the impact dynamics, probably because of the higher shear rate involved in such a phenomenon \cite{boyer16,zang14,zang13,ravi13,an12,laan14,german10,moon14,huh15}. The post impact spreading of a shear-thinning drop is faster than a Newtonian fluid having the same zero-shear viscosity and slower than a Newtonian fluid having the same infinite-shear viscosity. The subsequent recoil is slower than a Newtonian fluid with the same infinite-shear viscosity and is highly dependent both on viscosity and on surface wettability. Generally, spreading occurs at higher speed as if the fluid features a lower viscosity, while recoil dynamics is slower, as for a more viscous fluid \cite{ravi13,an12,german10,laan14,moon14,Seevaratnam07}. Normal stress fluids show negligible deviations from the Newtonian reference in the spreading phase, whereas bouncing and rebounding are highly suppressed and the receding contact line is retarded \cite{smith10,zang14,zang13,son09,smith14}. Shear-thickening fluids exhibit even more unusual behaviours, regarding both dynamics and the final state, in particular the maximum deformation does not depend on the velocity \cite{boyer16}.\\ 
The first paper \cite{Morita09} focusing on the sliding of non-Newtonian drops considers polystyrene/acetophenone solutions moving down homogeneous silicon-coated glass where sliding is often affected by the presence of pearling. In such a regime, drops are generally found to slide faster than the silicon oil chosen as Newtonian reference. However, the rheological properties of the investigated polymer solutions are not accurately detailed and the comparison involves set of data extracted from different papers published by different groups. 
%
%
Our previous work reports~\cite{EPJnoi} a joint experimental and numerical study which analyses the sliding of Xanthan (a stiff polymer) aqueous solutions featuring a pronounced shear-thinning viscosity.
%
We find that, at variance with Newtonian fluids where the relation between velocity and driving force is linear in the steady sliding regime, for viscoelastic drops this relation is sublinear and depends on the specific polymer and its concentration. 
Preliminary lattice Boltzmann (LB) simulations ascribe such a deviation to the presence of normal stresses developed in the drop during sliding. However, Xanthan solutions are often used as model system for power-law shear-thinning and only concentrated solutions exhibit elastic behaviours due to the entanglement between different stiff polymer chains \cite{Macosko78}.\\
To confirm the results of the LB calculations, we then decided to perform sliding experiments using drops of polyacrylamide (PAA) solutions with different molecular weights. These polymeric solutions exhibit a marked elastic behaviour and only marginal shear thinning. To highlight the elastic contributions we systematically compared the sliding of flexible polymer (PAA) drops with those made of stiff polymers (Xanthan). The experimental results were again complemented with LB simulations, with the specific purpose of investigating the effects of the elastic parameters (polymer relaxation time and finite extensibility).
Numerical simulations were also exploited to visualise the distribution of polymer feedback stresses inside the drop and the shape of the drop during sliding.\\ 
The paper is organised as follows: section \ref{methodology} deals with the experimental and numerical techniques applied in this study, the corresponding results are described in sections \ref{exp_res} and \ref{numerical_results}, while conclusions are reported in section \ref{conclusion}.

%
%
%
\section{Material and Methods} \label{methodology}
%
%
\subsection{Experiments} \label{sec:expmethods}
\begin{figure}[!htb]
\centering
\includegraphics[scale=1]{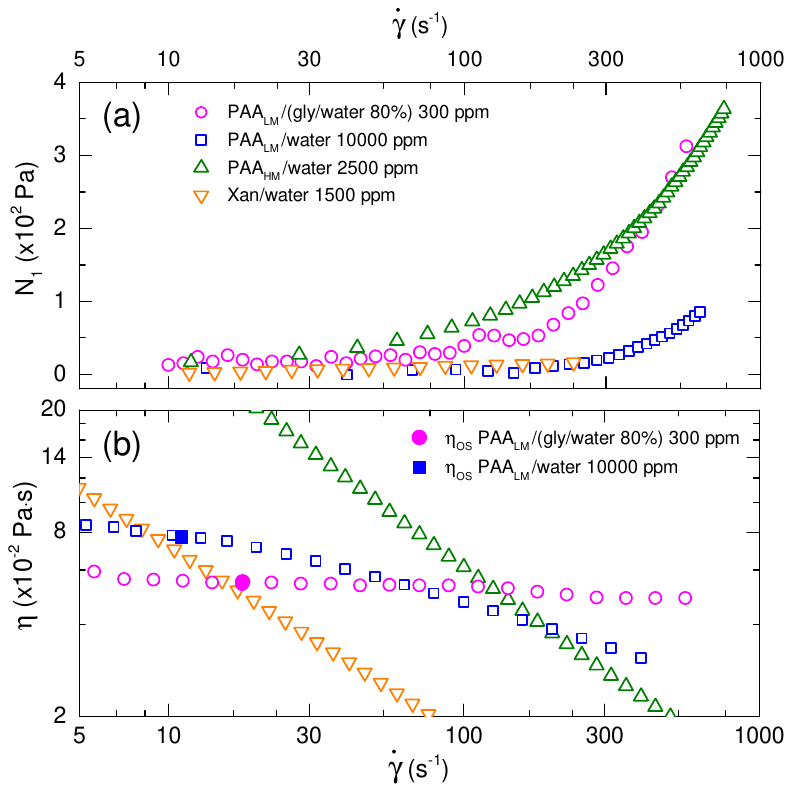}\\
\caption{Panel (a): first normal stress difference for flexible (PAA at different $M_w$) and stiff polymer (Xanthan) solutions as a function of the shear rate. Panel (b): shear dependent viscosity corresponding to the same solutions (same symbols) plotted in panel (a). Open circles show the shear independent viscosity of a Boger fluid obtained by suspending PAA$_{\textrm{LM}}$ in a water/glycerine mixture. Filled square and circle are the viscosity measured at fixed shear rate with a glass capillary viscometer for PAA$_{\textrm{LM}}$ in water and glycerol/water mixture, respectively. Xanthan data of both $N_1$ and $\eta$ are reproduced from~\cite{EPJnoi}, with kind permission of The European Physical Journal (EPJ).\\
}
\label{fig:visco}
\end{figure}

We considered polyacrylamide (PAA) featuring different molecular weights: $M_{w}\sim 10^{6}$ $g \cdot mol^{-1}$ (Sigma Aldrich) to which we will refer as PAA of ``low'' molecular weight, PAA$_\textrm{LM}$, and $M_w \sim 10^{7}$  $g \cdot mol^{-1}$ (Polysciences, Inc.) which we will label as PAA$_\textrm{HM}$.  We dispersed PAA polymers either in water or in a mixture of glycerol and water at a glycerol concentration of 80\% w/w. In both cases, solutions had concentrations ranging in the dilute or semi-dilute regime \cite{Rafai,Rafai05,Callaghan00} as listed in Table \ref{viscosity_table}.
The rheological properties of PAA solutions were probed through two different methods: i) measurements performed with a glass capillary (Ostwald viscometer) provided an evaluation of the viscosity $\eta_{\tiny\mbox{OS}}$ at fixed shear rate in the range $\dot{\gamma}\approx $ 1-20 $s^{-1}$ (see Table \ref{viscosity_table}); ii) using a plate-plate rheometer (Ares TA Instruments, New Castle, DE, USA) both the viscosity $\eta(\dot{\gamma})$ and the first normal stress difference $N_1(\dot{\gamma})$ were determined as a function of the shear rate ($N_1=\tau_{11}-\tau_{22}$, where $\tau_{11}$ and $\tau_{22}$ are the diagonal elements of the stress tensor~\cite{tanner}).
As shown in figure~\ref{fig:visco}-a, PAA solutions report $N_1$ increasing either with the shear rate or the molecular weight, in agreement with data reported for similar solutions \cite{Arratia}. Indeed, at the highest shear rates, $N_1(\dot{\gamma})$ of PAA$_\textrm{HM}$ is about four times greater than the one of PAA$_\textrm{LM}$.   
In parallel, from Fig.~\ref{fig:visco}-b it results that, while the viscosity of water solutions of PAA$_\textrm{LM}$ is nearly shear independent over a wide range of polymer concentration, PAA$_\textrm{HM}$/water clearly exhibits a power-law behaviour.
In fact, at concentrations below 2500 ppm, the viscosity of PAA$_{\textrm{LM}}$/water does not depend on the shear rate, in agreement with \cite{Rafai,Arratia}. As the polymer concentration is increased $\eta$ becomes weakly shear dependent. At 10000 ppm the solution of PAA$_{\textrm{LM}}$/water behaves as weak shear thinning fluid, being $\eta(\dot{\gamma})$ decreasing of about 50$\%$ over the explored range of shear rates. However, this thinning is by far much lower than the one commonly observed for power law fluids.
To complete this analysis, we also report for comparison the behaviour of either stiff polymers (Xanthan solutions, used in similar sliding experiments \cite{EPJnoi}) or Boger fluids (i.e. strictly shear independent viscosity~\cite{derzsi2013flow}) made of PAA$_{\textrm{LM}}$ dispersed at 300 ppm in a mixture glycerol/water 80\% (Fig.~\ref{fig:visco}-a,b). The corresponding $\eta_{\tiny\mbox{OS}}$ and $\eta(\dot{\gamma})$ taken at low shear rates are in good agreement with each other.
We point out that PAA$_\textrm{HM}$ and the Boger fluids show similar elastic properties, while the thinning behaviour is totally different. The same applies to Xanthan 1500~ppm and PAA$_\textrm{LM}$ 10000~ppm water solutions.\\
%
%
%
%
\begin{table}[!htb]
\small
  \caption{Viscosity of the non-Newtonian PAA$_{\textrm{LM}}$ solutions, $\eta_{\tiny\mbox{OS}}$, measured with a glass capillary (Ostwald) viscometer.} 
  \label{viscosity_table}
\begin{tabular*}{0.48\textwidth}{@{\extracolsep{\fill}}ccc}
\hline
Liquid & concentration   & $\eta_{\tiny\mbox{OS}}$ \\ 
& (ppm w/w)   & (mPa$\cdot$s)  \\ \hline
PAA$_{\textrm{LM}}$/water & 250 &  1.09$\pm$0.03 \\ 
PAA$_{\textrm{LM}}$/water & 1500 & 2.09$\pm$0.06 \\ 
PAA$_{\textrm{LM}}$/water & 2500 &   4.0$\pm$0.1 \\ 
PAA$_{\textrm{LM}}$/water & 5000 &   9.1$\pm$0.3 \\ 
PAA$_{\textrm{LM}}$/water & 10000 &   77$\pm$2 \\ 
PAA$_{\textrm{LM}}$/(glycerol/water 80\%) & 300 & 55$\pm$2 \\\hline
\end{tabular*}
\end{table}
%

The substrate used for sliding experiments was a homogeneous, polycarbonate (PC) plate, whose wettability properties were determined using the sessile drop method \cite{Toth11}. The sample was characterised both by static and dynamic contact angles (CA) for each polymer solution. Values were similar for all solutions and their average values were: static CA $\theta_s=(84\pm4)^\circ$, advancing CA $\theta_{a}=(88\pm5)^\circ$ and receding CA $\theta_{r}=(63\pm4)^\circ$. Sliding measurements of 30 $\mu \text{L}$ drops were performed similarly to \cite{PRLnoi,PREnoi,Langmuir,EPJnoi}. Briefly, the desired liquid was first deposited on the already inclined sample and the drop motion was followed with a camera. The drop contour was then identified with a custom-made program, and the velocity of the steady-state motion was obtained by the temporal evolution of the frontal contact point \cite{PRLnoi,PREnoi,EPJnoi}.
\newline
The appropriate dimensionless numbers required to compare the dynamics of drops of different fluids are suggested by the analysis of the sliding problem for Newtonian drops \cite{Podgorski01,Kimetal02,Legrand05}. Specifically, they are the \textit{Capillary number} 
\begin{equation}
\mbox{Ca}=\frac{\eta U}{\sigma}
\end{equation}
computed from the steady velocity $U$, surface tension $\sigma$ and viscosity $\eta$, and the \textit{Bond number} 
\begin{equation}
\mbox{Bo}=V^{2/3}\rho g\sin\alpha/\sigma
\end{equation} 
where $V$ is the drop volume, $\rho$ the fluid density, $g$ the gravity acceleration and $\alpha$ the inclination angle of the plane. 
The balance of the forces acting on a Newtonian drop can be written as
\begin{equation}
\mathrm{{Ca}\propto {Bo}-{Bo_{c}}}  \label{eq:scaling}
\end{equation}%
where ${Bo}_{c}=V^{2/3}\rho g\sin \alpha _{c}/\sigma$ is the critical Bond number, below which the drop remains pinned and does not move~\cite{Furmidge62}.


\subsection{Numerical Simulations}\label{sec:methodsnumerics} 

To perform numerical simulations of polymeric fluids we used a Navier-Stokes (NS) description for the solvent coupled with constitutive equations for the stress tensor accounting for the (coarse grained) effects of polymer molecules. We adopted the FENE-P constitutive model, which hinges on a pre-averaging approximation applied to an ensemble of non interacting Finitely Extensible Nonlinear Elastic (FENE) dumbbells~\cite{birdpaper,bird87,Herrchen97,Wagner05,Lindner03,Arratia}. The NS hydrodynamic equations were reproduced with the help of lattice Boltzmann (LB) simulations of sliding drops \cite{Moradi,PREnoi,Kusumaatmaja}. The specific LB model used was validated in dedicated papers by some of the authors~\cite{SbragagliaGuptaScagliarini,SbragagliaGupta,EPJnoi} and we refer the interested reader to such papers for the technical details on LB. We only recall here the reference continuum equations. Inside the drop we solve the NS equations for the hydrodynamic velocity ${\bm u}_{d}$, coupled to the FENE-P equations for the polymer conformation tensor ${\bm {\mathcal C}}$~\cite{birdpaper,bird87,Herrchen97}:
\begin{eqnarray}  \label{EQ} \nonumber 
\rho _{d}\left[ \partial _{t}\bm u_{d}+({\bm u}_{d}\cdot {\bm\nabla })\bm %
u_{d}\right] &=&-{\bm\nabla }P_{d}+{\bm\nabla }\left( \eta _{d}({\bm\nabla }{%
\bm u}_{d}+({\bm\nabla }{\bm u}_{d})^{T})\right)+\\%
 &+&\frac{\eta _{P}}{\tau _{P}}%
{\bm\nabla }\cdot \lbrack f(r_{P}){\bm {\mathcal C}}]+{\bm g}\rho _{d}\sin
\alpha  \label{NSb} \nonumber \\
\nonumber \partial _{t}{\bm {\mathcal C}}+(\bm u_{d}\cdot {\bm\nabla }){\bm {\mathcal
C}} &=&{\bm {\mathcal C}}\cdot ({\bm\nabla }{\bm u}_{d})+{({\bm\nabla }{\bm u%
}_{d})^{T}}\cdot {\bm {\mathcal C}}+\\%
\nonumber &-&\frac{{f(r_{P}){\bm {\mathcal C}}}-{\bm I%
}}{\tau _{P}} \label{FENEb}
\end{eqnarray}
where $P_{d}$ is the drop bulk pressure and $({\bm\nabla }{\bm u}_{d})^{T}=\partial_j u_{d,i}$ is the transpose of the tensor $\partial_i u_{d,j}$. The fluid stress tensor in the NS equations is the sum of two contributions: the viscous stress ($\eta _{d}({\bm\nabla }{\bm u}_{d}+({\bm\nabla }{\bm u}_{d})^{T})$) and the polymer feedback stress ($\frac{\eta _{P}}{\tau _{P}} f(r_{P}) {\mathcal C}$). The body force ${\bm g}\rho _{d}\sin \alpha $ is applied in a given direction to mimic the effect of gravity down the inclined plane. The polymer-conformation tensor ${\bm {\mathcal C}}$ represents the (ensemble) average of the second order tensor constructed with the bead-to-bead separation vector~\cite{birdpaper,bird87,Herrchen97}. It gives information about the stretching and orientation of the polymers in the flow. The characteristic polymer time $\tau _{P}$ regulates the relaxation of the polymer conformation tensor towards the unstretched (equilibrium) value, ${\bm {\mathcal C}}={\bm I}$. Moreover, FENE polymers can only be stretched up to a finite amount, which is parametrised by a maximum extensional length squared $L^{2}$. The FENE-P potential $f(r_{P})$~\cite{birdpaper,bird87,Herrchen97} ensures such finite extensibility ($r_p=Tr({\bm {\mathcal C}})$). In the outer continuous phase we integrated the NS equations for the velocity ${\bm u_{c}}$ 
\[
\rho _{c}\left[ \partial _{t}\bm u_{c}+({\bm u}_{c}\cdot {\bm\nabla })\bm %
u_{c}\right] =-{\bm\nabla }P_{c}+{\bm\nabla }\left( \eta _{c}({\bm\nabla }{%
\bm u}_{c}+({\bm\nabla }{\bm u}_{c})^{T})\right)
\]
In all numerical simulations, we made use of a neutral wetting boundary condition ($\theta_s=90^\circ$) and a Neumann boundary condition for the conformation tensor in contact with the flat substrate. The shear viscosity of the drop is the sum of two contributions, coming from the polymers ($\eta _{P}$) and the solvent ($\eta _{d}$). With respect to the outer phase, the viscosity ratio $\eta _{c}/(\eta _{d}+\eta _{P})$ is kept fixed to unity. This allows us to analyse the behaviour of Newtonian drops ($\eta_P=0$) in a Newtonian outer phase, and to compare it with the corresponding non-Newtonian ($\eta_P \neq 0$) problem, for the same shear viscosity and viscosity ratio~\cite{SbragagliaGuptaScagliarini,SbragagliaGupta,EPJnoi}. We remark that the very same model was used to highlight the importance of normal stresses in the sliding drop problem~\cite{EPJnoi}. This was done for a specific (i.e. fixed $\tau_P$ and $L^2$) realisation of the FENE-P parameters. The focus here is on the quantitative impact of a change of these two parameters on the relation between the Bond number and the Capillary number.


\section{Experimental Results} \label{exp_res}

\begin{figure}[!htb]
\centering
\includegraphics[scale=1]{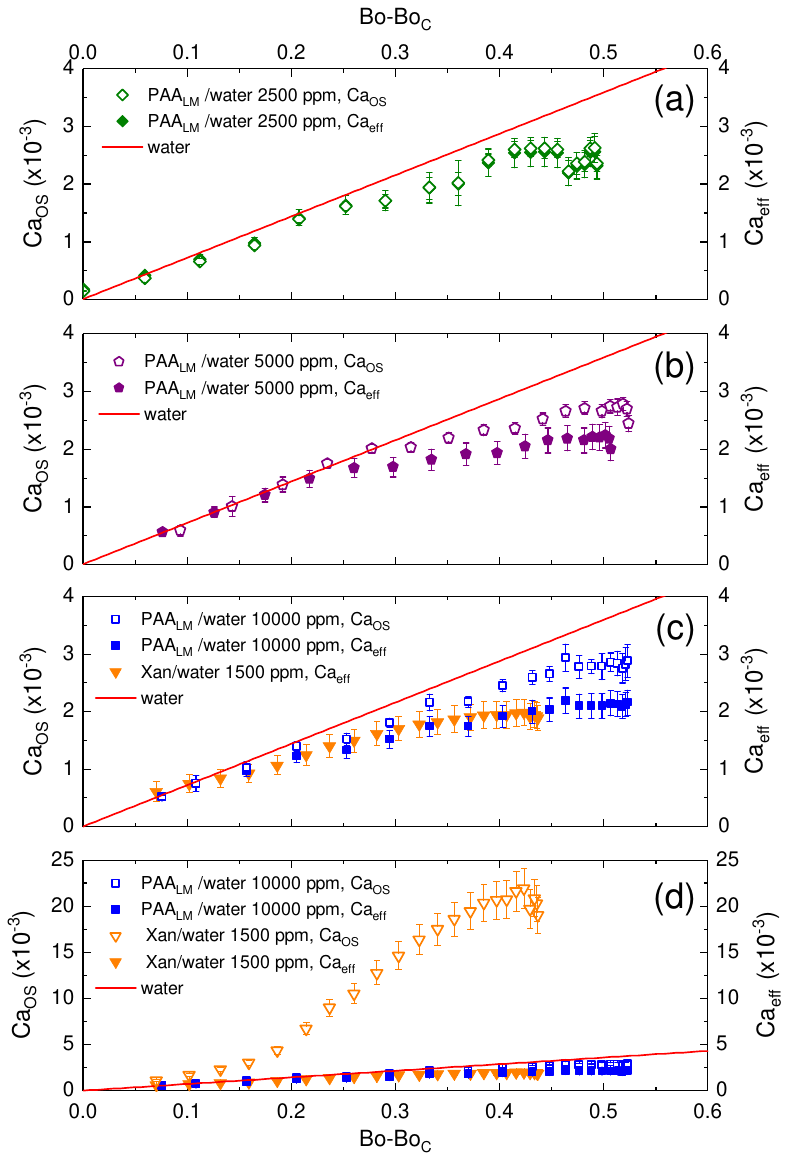}
\caption{Capillary number as a function of the Bond number for sliding drops made of $\mbox{PAA}_{\tiny{LM}}$/water at 2500 ppm (a), 5000 ppm (b), 10000 ppm (c) and Xanthan/water at 1500 ppm in comparison to $\mbox{PAA}_{\tiny{LM}}$/water 10000 ppm (c,d). Left vertical axis displays the shear independent Capillary number $\mbox{Ca}_{\tiny\mbox{OS}}$ (Eq.~\ref{eq:Caos}). 
Right vertical axis refers to the shear dependent Capillary number $\mbox{Ca}_{\tiny\mbox{eff}}$ (Eq.~\ref{eq:caeff}). Bars are standard deviation over different measurements. Line in panels (a,b,c,d) is the linear fit of water data used as Newtonian reference. Xanthan and water data are reproduced from~\cite{EPJnoi}, with kind permission of The European Physical Journal (EPJ).}
\label{fig:3marie}
\end{figure}

In Fig.~\ref{fig:3marie} we report results from the sliding experiments performed with PAA$_{\textrm{LM}}$/water solutions at various concentrations. We show the dependence of the Capillary number on the Bond number, a common method to assess the force balance between interfacial and bulk forces acting on the drop~\cite{Kimetal02,Legrand05,Morita09,Podgorski01}. 
To compute the Capillary number of the polymer solutions we first neglected the dependence of the viscosity from the shear rate. For  $\mathrm{PAA_{LM}}$ with concentration lower than 5000 ppm this is a reasonable approximation according to the rheology discussed in Fig.~\ref{fig:visco}. We therefore define $\mbox{Ca}_{\tiny\mbox{OS}}$ as:
\begin{equation}\label{eq:Caos}
\mbox{Ca}_{\tiny\mbox{OS}}=\frac{\eta_{\tiny\mbox{OS}} U}{\sigma}
\end{equation}
where $\eta_{\tiny\mbox{OS}}$ is the  viscosity measured at a fixed shear rate with a glass capillary viscometer, reported in Table \ref{viscosity_table}. Data of $\mathrm{PAA_{LM}}$ corresponding to low inclinations follow the Newtonian linear trend, while at higher $\mathrm{{Bo}-{Bo}_{c}}$ the Ca number stops increasing linearly and shows a sublinear behaviour, with a saturation in the Ca number at larger concentrations. These findings are similar to what we have reported for stiff polymers ~\cite{EPJnoi}.
However, we point out that the present analysis did not include the shear rate dependence of the $\mbox{Ca}$ number. To keep into account for (weak) shear-thinning effects
we followed the same approach that we have introduced for Xanthan solutions in our previous study~\cite{EPJnoi}. We defined an effective shear rate $\dot{\gamma}_{\tiny\mbox{eff}}=U/\lambda$, $\lambda$ being a characteristic length scale phenomenologically introduced to account for an ``effective gradient'' inside the drop, and $U=U(\alpha)$ the steady sliding velocity corresponding to the sliding angle $\alpha$. The resulting Capillary number was therefore labelled as $\mbox{Ca}_{\tiny\mbox{eff}}$ accordingly to the definition:
\begin{equation}\label{eq:caeff}
\mbox{Ca}_{\tiny\mbox{eff}}=\frac{\eta(\dot{\gamma}_{\tiny\mbox{eff}}) U}{\sigma}.
\end{equation}
Fig.~\ref{fig:3marie}-a,b,c shows that $\mbox{Ca}_{\tiny\mbox{eff}}$  is always smaller than 
the shear independent  $\mbox{Ca}_{\tiny\mbox{OS}}$. This decrease is proportional to the polymers concentration and it is roughly of the same entity of the decrease of the viscosity discussed in Fig.~\ref{fig:visco}.
By using $\mbox{Ca}_{\tiny\mbox{eff}}$ we were then able to compare the sliding of viscoelastic drops made of polymer's solutions featuring similar $N_1$ but different shear thinning $d\eta/d\dot{\gamma}$ (see Fig.~\ref{fig:3marie}-c,d). 
The role played by the elasticity to produce a sublinear trend in the relation~\eqref{eq:scaling} during sliding has already been addressed in~\cite{EPJnoi}. In the case of Xanthan, the elasticity is attributed to the network formed by the entanglement of stiff chains~\cite{Macosko78}, for PAA it originates from the intrinsic flexibility of single polymer chains.
%
%
%
In  Fig.~\ref{fig:3marie}-c, we directly compare  Xanthan/water 1500 ppm and $\mbox{PAA}_{\tiny{LM}}$/water 10000 ppm and find that the data overlap nicely when plotted in terms of $\mbox{Ca}_{\tiny\mbox{eff}}$.
We point out that in Fig.~\ref{fig:3marie}-c the Xanthan and $\mbox{PAA}_{\tiny{LM}}$ data do not cover the same range, because the contact angle hysteresis (e.g.the $\mbox{Bo}_c$) was different in the two cases.
%
%
Figure~\ref{fig:3marie}-d clearly shows that, while the use of a constant viscosity introduces  just small corrections for PAA$_{\textrm{LM}}$/water, in the case of power-law fluids like Xanthan/water solutions $\mbox{Ca}_{\tiny\mbox{eff}}$ and $\mbox{Ca}_{\tiny\mbox{OS}}$ result overturned even respect to water, confirming that the comparison of the sliding behaviour of drops made of different polymers can be properly accomplished only by using $\mbox{Ca}_{\tiny\mbox{eff}}$~\cite{EPJnoi}.
In the case of Xanthan we computed $\mbox{Ca}_{\tiny\mbox{OS}}$ by considering  the $\eta_{\tiny\mbox{OS}}$ measured for $\mbox{PAA}_{\tiny{LM}}$/water 10000 ppm. Indeed,  this value nicely approximates the (common) viscosity resulting from the intersection between the rheological curves of Xanthan/water 1500 ppm and $\mbox{PAA}_{\tiny{LM}}$/water 10000 ppm.\\
As explained in~\cite{EPJnoi} the sublinear portion of the Ca {\it vs.} Bo curve is the result of a supplementary contribution to the balance of the forces acting on the sliding drop, provided by the presence of the polymers. 
We checked whether this contribution determines in a morphological change of the sliding drops~\cite{Seevaratnam05,Seevaratnam07}. 

\begin{figure}[!htb]
\centering
\includegraphics[width=\columnwidth]{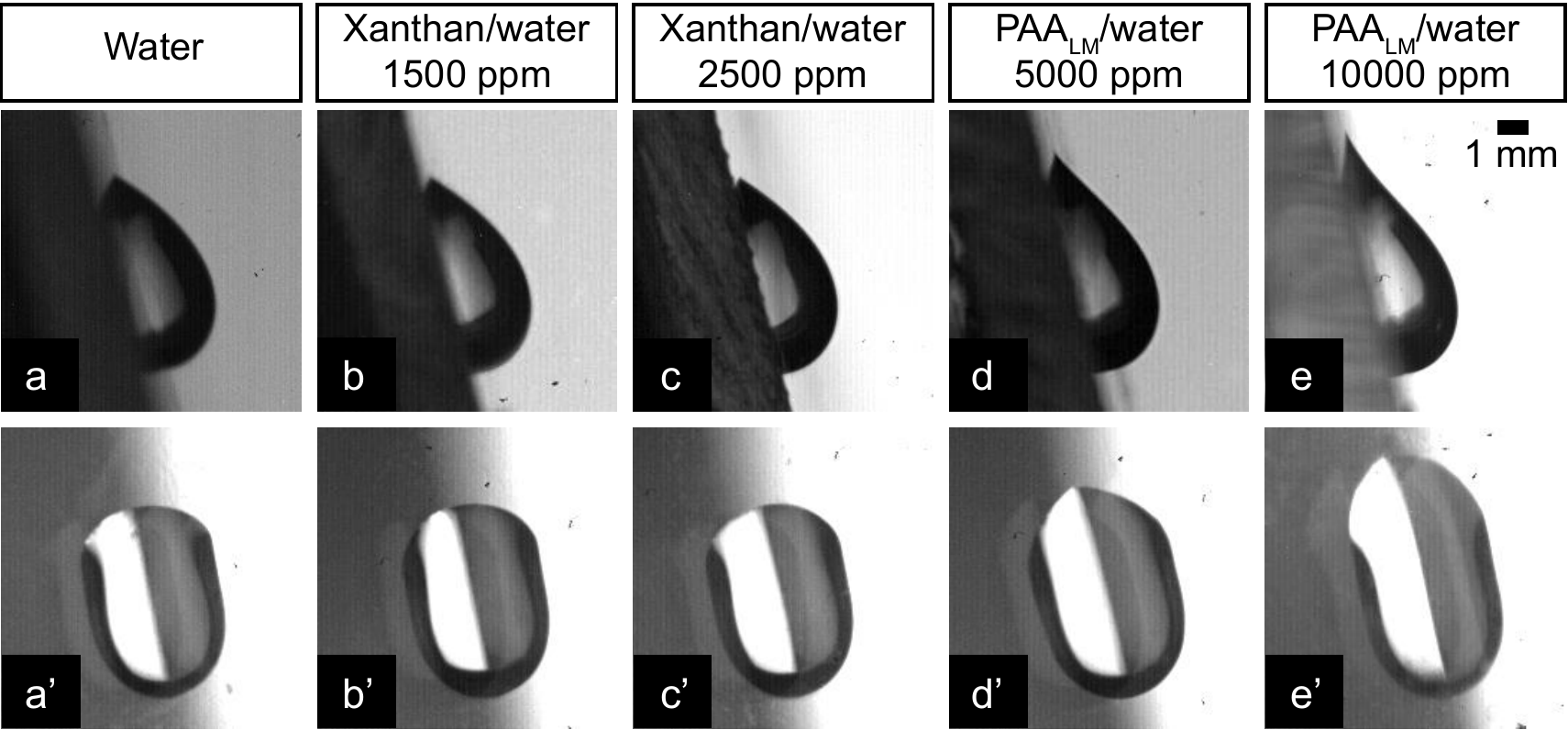}
\caption{Snapshots showing the side (a-e) and bottom (a'-e') view of 30 $\mu\textrm{L}$ drops sliding down a homogeneous PC surface inclined by the same angle ($\alpha=76^{\circ}$).
}
\label{fig:snaps}
\end{figure}

To this aim, we imaged the drops motion in the steady sliding regime, as shown in some examples reported in Fig.~\ref{fig:snaps} at inclinations corresponding to $\mbox{Bo} -\mbox{Bo}_c \simeq 0.4$. Both the side view (pictures a-e) and the bottom view (pictures a'-e') suggest that, for a fixed plane inclination $\mbox{PAA}_{\tiny{LM}}$, drops get elongated and develop a cornered rear tip, as observed for Newtonian viscous oils~\cite{Podgorski01}. In addition, drop stretching seems to become more pronounced as the PAA concentration increases. 
At similar Capillary numbers Newtonian drops do not exhibit such a remarkable stretching~\cite{Legrand05}. Therefore, it seems plausible to attribute the observed behaviour to the presence of the $\mbox{PAA}_{\tiny{LM}}$ polymers.
Although the drop shape is highly sensitive to surface features such as wettability, homogeneity and cleanliness~\cite{sun05}, in the case of Xanthan drops no appreciable stretching is observed (see Fig~\ref{fig:snaps},b,b'), despite the $\mbox{Bo}$ dependence of the effective $\mbox{Ca}$ number is the same as for $\mbox{PAA}_{\tiny{LM}}$ drops (see Fig.~\ref{fig:snaps}-d, d' and Fig.~\ref{fig:snaps}-e, e').
%
This different behaviour may be due to different shear thinning. In fact, the side view indicates that the morphological changes mainly occur in the drop's wedge, close to the rear contact line, where the shear rate is maximum.
\\
%
%
%

\begin{figure}[!htb]
\centering
\includegraphics[scale=1]{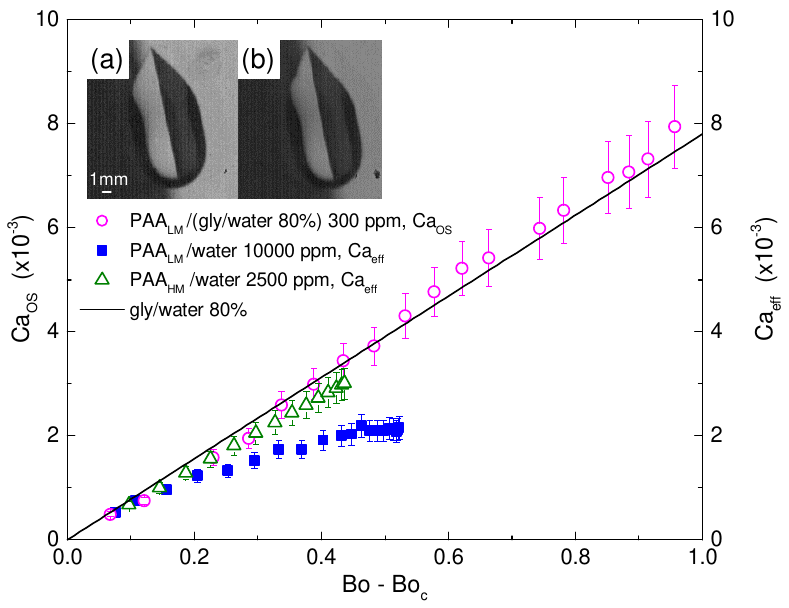}
\caption{Dimensionless $\mathrm{Ca}_{\tiny\mbox{OS}}$ (left vertical axis) and $\mathrm{Ca}_{\tiny\mbox{eff}}$ (right vertical axis) as a function of $\mathrm{{Bo}-{Bo}_{c}}$ for the Boger fluid in comparison to PAA$_{\textrm{LM}}$/water at 10000 ppm and PAA$_{\textrm{HM}}$/water at 2500 ppm. Line is the linear fit to glycerol/water 80\% w/w (the Boger solvent, data not reported) according to Eq.~\ref{eq:scaling}. Bars are standard deviation over different measurements. Pictures taken from the bottom view of the experimental setup show drops sliding down a homogeneous PC surface inclined by the same angle ($\alpha=76^{\circ}$): glycerol/water 80\% w/w (a) and the Boger fluid (b).\\
}
\label{fig:exp_PAA_glycerol}
\end{figure}

%
%
%
%
%
To better understand the contribution of a large $N_1$ to the drop dynamics, we dispersed $PAA_{LM}$ in a mixture of glycerol/water at 80\% w/w  obtaining the Boger fluid described in Sec.~\ref{sec:expmethods}
and we used higher molecular weight $\mbox{PAA}_{\tiny{HM}}$ in water. Overall, we performed sliding experiments with two polymer solutions featuring similar $N_1$, about four times larger than the one of both $PAA_{LM}$/water and Xanthan/water solutions (see Fig.~\ref{fig:visco}-a).  While the viscosity of $\mbox{PAA}_{\tiny{HM}}$/water is described by a power-law fluid model similar to Xanthan/water,  the viscosity of the Boger fluid is strictly shear independent by definition (see Fig.~\ref{fig:visco}-b).
In Fig.~\ref{fig:exp_PAA_glycerol} we report the $\mbox{Ca}$ \textit{vs.} $\mbox{Bo}$ curves of the sliding experiments, accompanied by characteristic snapshots of the drops taken at the same inclination as the snapshots reported in Fig.\ref{fig:snaps}. A comparison with the sliding of 10000 ppm $\mbox{PAA}_{\tiny{LM}}$/water, presented in Fig.~\ref{fig:3marie}-c, is also reported. The Newtonian reference for the Boger fluid is the mixture of glycerol/water 80\% w/w. As expected, the sliding of Newtonian drops made of this mixture yields a linear relation between $\mbox{Ca}$ and $\mbox{Bo}$, with a slope similar to the one of water drops.
Surprisingly, the same linear trend is also displayed by the Boger fluid in spite of the fact that its elastic effect ($N_1$) is more pronounced than the 10000 ppm of PAA$_{\textrm{LM}}$ in water (reported in Fig.~\ref{fig:exp_PAA_glycerol} for comparison).
This trend was tested over an extended range of both $\mbox{Ca}$ and $\mbox{Bo}$ numbers.
At high Ca the error bars get larger since the sliding was affected by the detachment of satellite microdrops from  the rear of the sliding drops (pearling~\cite{Legrand05}). 
In addition, sliding drops containing polymers (both $\mbox{PAA}_{\tiny{HM}}$ and $\mbox{PAA}_{\tiny{LM}}$) did not report any appreciable stretching with respect to the ones made by the corresponding Newtonian solvent, either water or glycerol/water mixture,
as shown in Fig.~\ref{fig:exp_PAA_glycerol}-a (glycerol/water mixture) and Fig.~\ref{fig:exp_PAA_glycerol}-b (Boger fluid).
These experimental observation suggest that the sublinear dependence in the $\mbox{Ca}$ {\it vs.} $\mbox{Bo}$ curve appears only for a limited range of $N_1$ values. Outside this interval the dynamical curves are linear and there is no extra stretching due to the presence of the polymer.
%
%
%
This echoes the experimental findings of Garoff and co-workers on dynamic wetting \cite{WeiGaroff07,WeiGaroff09} performed at fixed velocity (i.e. fixed $\mbox{Ca}$ number). In these investigations the authors found that the use of high molecular weight polymers did not introduce an added force at the contact line, compared to the intrinsic elasticity of the solvent~\cite{WeiGaroff07,WeiGaroff09}. The experimental findings reported in Fig.~\ref{fig:exp_PAA_glycerol} will be further discussed and unraveled in Sec.~\ref{numerical_results}  in view of the numerical simulations.\\
%
%
%


\section{Numerical Results} \label{numerical_results}

\begin{figure}[!htb]
\centering
\includegraphics[scale=1]{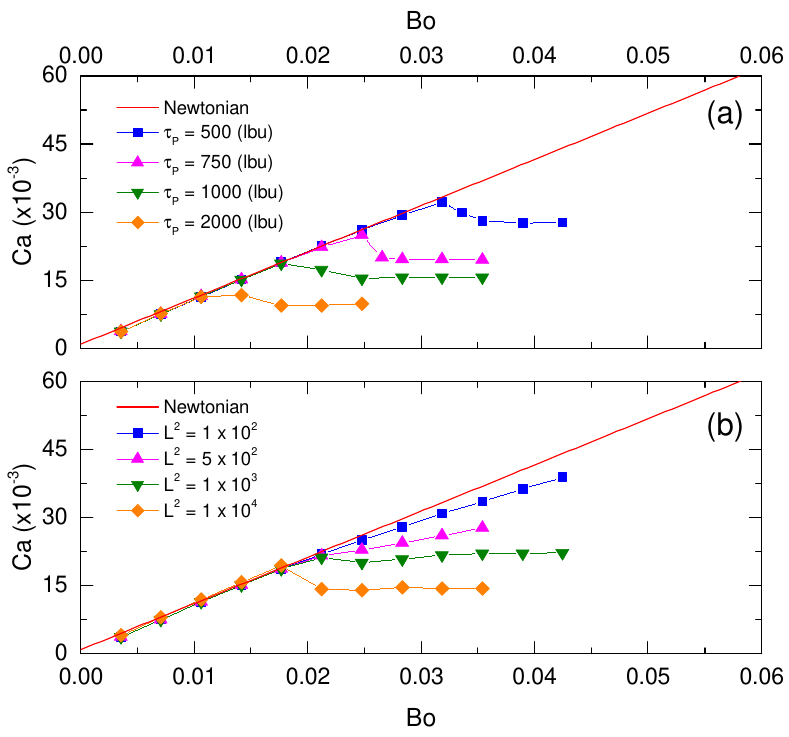}\\
\caption{$\mathrm{Ca}$ \textit{vs.} $\mathrm{Bo}$  curves for different values of $\protect\tau_P$, $L^2$ and $\protect\beta=\protect\eta_P/(\protect\eta_P+\protect\eta_d)$ in the FENE-P model (see section \ref{sec:methodsnumerics}). The line is a best fit based on Eq.~\ref{eq:scaling} for pure Newtonian data. Symbols are the results of the LB numerical simulations with viscoelastic drop (full lines are connections to the symbols as a guide for the eyes). Panel (a): we change $\protect\tau_P$ at fixed $L^2=5 \times 10^3$ and $\protect\beta=\protect\eta_P/(\protect\eta_P+\protect\eta_d)=0.2$. Panel (b): we change $L^2$ at fixed $\protect\tau_P=10^3$ (lbu), $\protect\beta=\protect\eta_P/(\protect\eta_P+\protect\eta_d)=0.2$. Data about Newtonian sliding are reproduced from~\cite{EPJnoi}, with kind permission of The European Physical Journal (EPJ).}
\label{fig:2}
\end{figure}

\begin{figure*}[!htbp]
\centering
\makeatletter
\par
\makeatother {\ \includegraphics[scale=0.15]{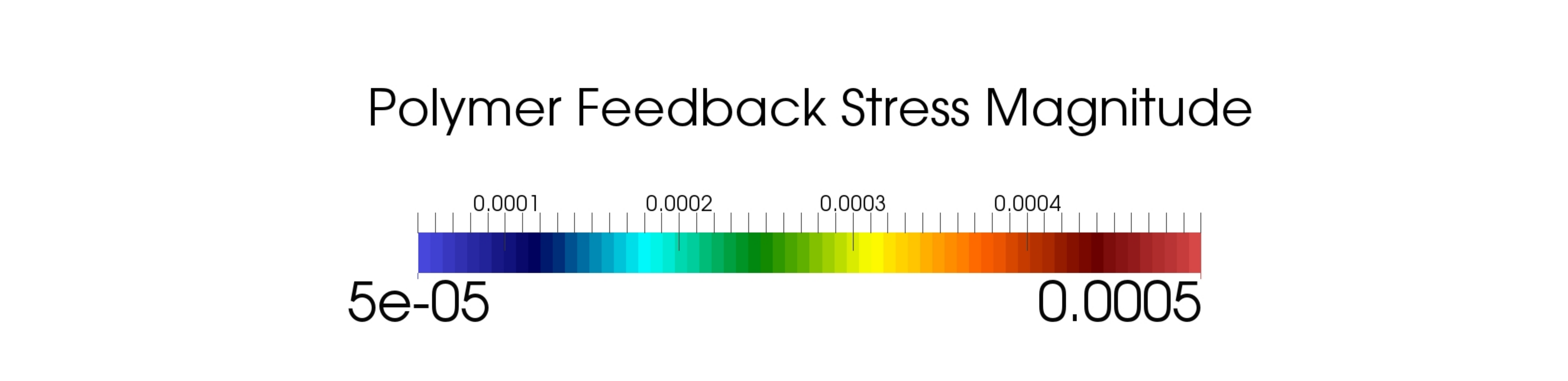}\\
}
\subfigure[\,\,Bo=0.024; $\tau_P=500$ (lbu), $L^2= 5 \times 10^3$] {\
\includegraphics[scale=0.1]{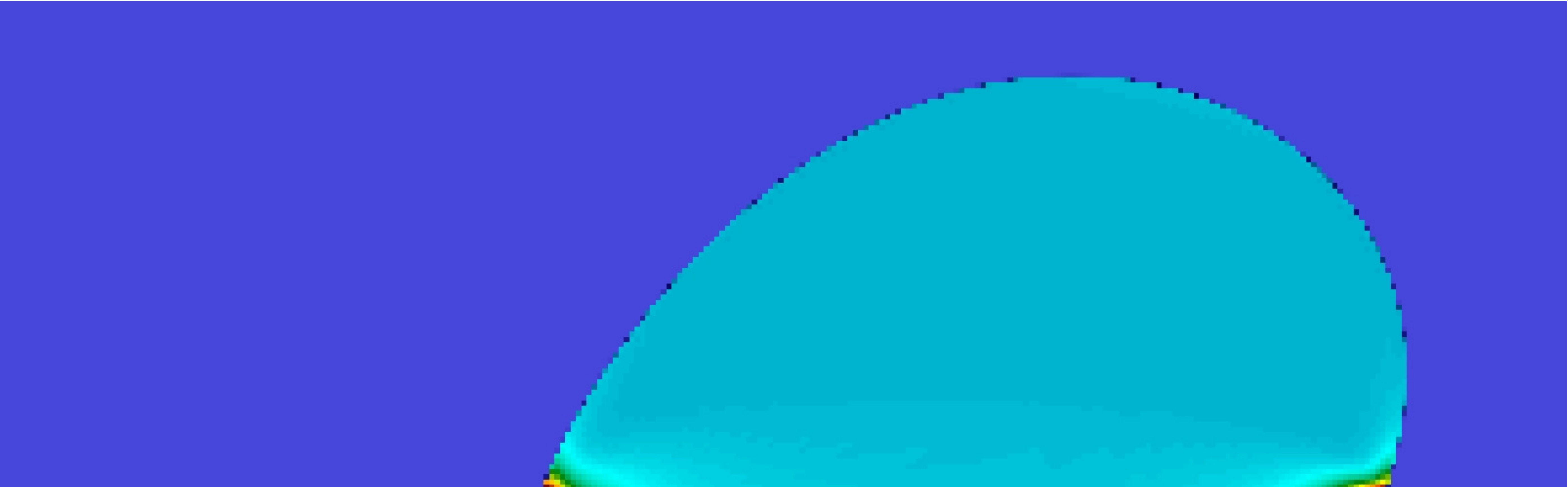} }
\subfigure[\,\, Bo=0.024; $\tau_P=1000$ (lbu), $L^2= 5 \times 10^3$] {\
\includegraphics[scale=0.1]{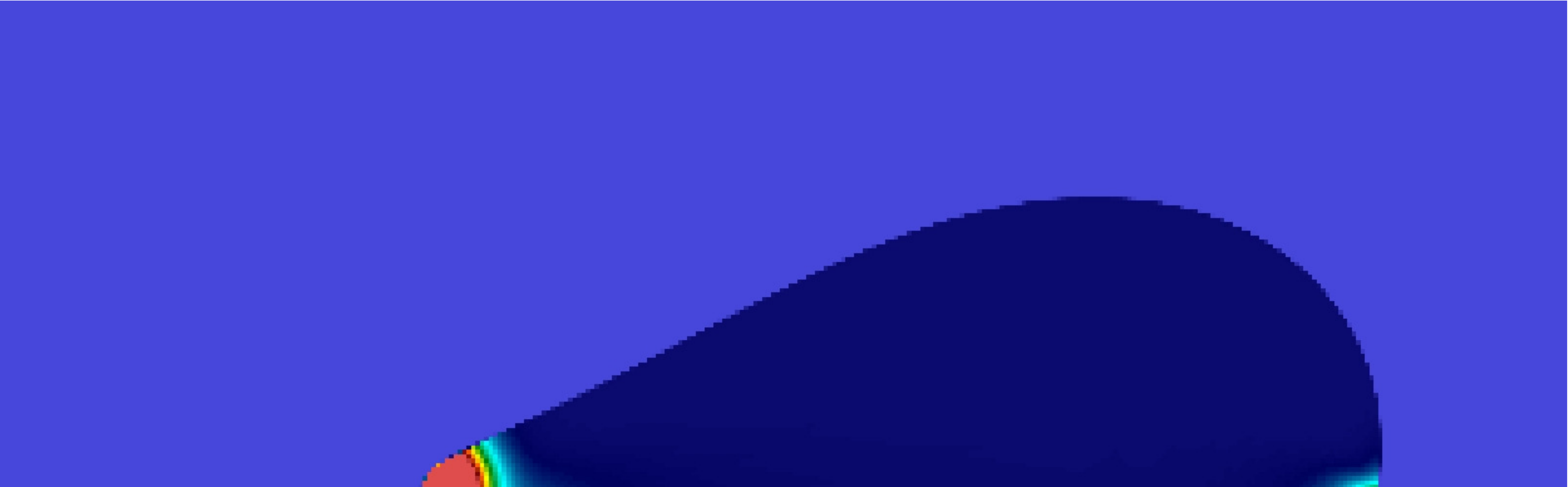} }
\subfigure[\,\, Bo=0.024; $\tau_P=2000$ (lbu), $L^2= 5 \times 10^3$] {\
\includegraphics[scale=0.1]{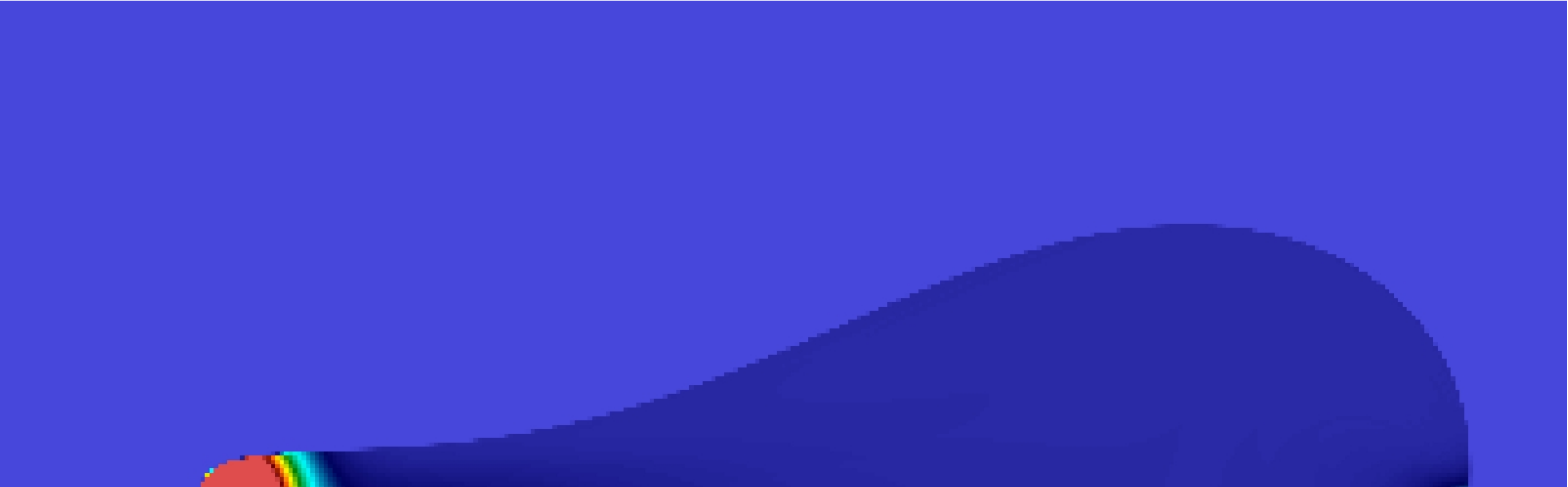} }
\caption{Stationary drop shapes and polymer feedback stresses inside the drop obtained for $L^2= 5 \times 10^3$ and $\mathrm{{Bo}=0.024}$ by varying the polymer relaxation time $\protect\tau_P$. Viscoelastic effects are pronounced in the wedge flow close to the contact line, while the bulk of the drop behaves essentially as a Newtonian fluid. For large $\protect\tau_P$, the emergence of viscoelastic stresses in the wedge flow region causes an extra bending of the non-ideal interface near a moving contact line and the drop gets elongated.}
\label{fig:3}
\end{figure*}

\begin{figure}[htb]
\centering
\includegraphics[scale=1]{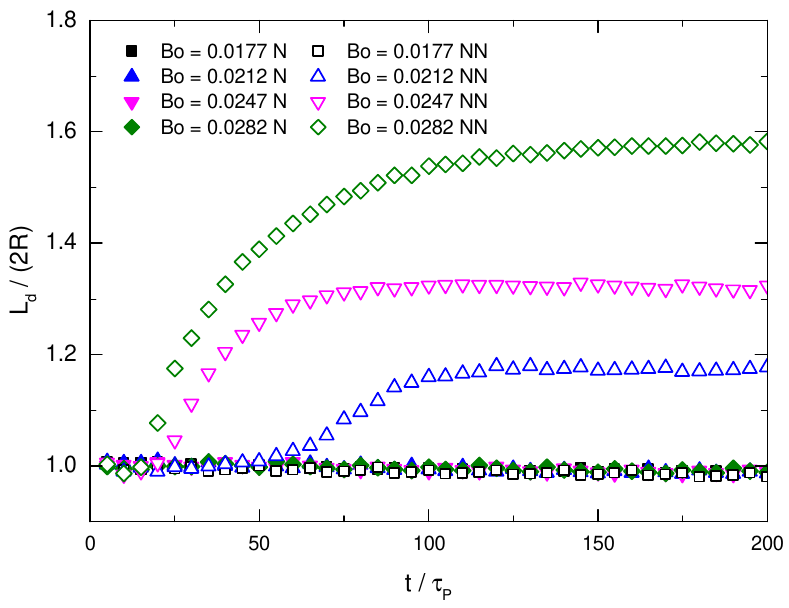}
\caption{Time evolution of the drop length $L_d$ normalised to the drop diameter $2R$ at rest for Newtonian (N, filled symbols) and non-Newtonian (NN, open symbols) fluids and different Bo numbers. Newtonian and non-Newtonian drops sliding at $\mathrm{Bo}\leq 0.0177$ have the same length (data not reported). In all cases we used a fixed $L^2= 5 \times 10^3$ and polymer relaxation time $\protect\tau_P=10^3$ (lbu). Notice that all filled symbols (Newtonian drops) well collapse on the same $L_d$, confirming that the elongation of the drop is a distinctive feature of viscoelasticity for the $\mathrm{Bo}$ considered.}
\label{fig:4}
\end{figure}

To complement the experimental data we have considered idealised models where viscoelastic effects can be tuned and visualised. 
Even though the fluids we have considered in the numerical model differ from those of the real experiments, numerical simulations are quite useful to reveal the importance of viscoelastic stresses, including their distribution, which cannot be deduced from the experiment. We refrain from establishing a deeper quantitative connection between numerical simulations and experiments, rather we use numerical simulations to explore the effects of normal stresses in an idealised fluid where thinning effects are minimised. Specifically, for the problem of a FENE-P drop, we extended our previous investigations \cite{EPJnoi} and addressed systematically the importance of the free parameters in the model (mainly $\tau_{P}$ and $L^{2}$, see section \ref{sec:methodsnumerics}) in promoting the emergence of sublinear (and plateau) behaviours in the $\mathrm{Ca}$ \textit{vs.} $\mathrm{{Bo}}$ curve. We could also visualise the distribution of the polymer feedback stresses during the motion of the drop, thus correlating the distribution of those stresses to the interface shape and the resulting macroscopic velocity.\\
Echoing the experimental results shown in section~\ref{exp_res}, we report in Fig.~\ref{fig:2} the $\mathrm{Ca}$ \textit{vs.} $\mathrm{Bo}$ curve extracted from the steady state drop velocity for various $\tau_P$ and $L^2$, at fixed $\beta=\eta_P/(\eta_d+\eta_P)=0.2$. Fig.~\ref{fig:2}-a considers a fixed $L^2=5\times 10^3$ with various $\tau_P$, the latter expressed in computational lattice Boltzmann units (lbu). An important point of discussion emerges on the role of the parameter $\tau_P$ in the FENE-P model. One expects viscoelastic effects to matter when the polymer relaxation time $\tau_P$ is of the order of the convective time scale~\cite{Yueetal12}, $\tau_{\tiny\mbox{fluid}} \approx \delta/U$, set by the characteristic velocity $U$ and the characteristic length $\delta$ of the problem. For the sliding drop, the characteristic velocity is taken from the sliding velocity, while we expect $\delta$ to be of the order of few interfaces widths for the diffuse interface methods \cite{Yueetal12}. For our numerical simulations we could sustain stable computations for a range of relaxation times $\tau_P$ such that $\tau_P/\tau_{\tiny\mbox{fluid}}$ varies from values much below $1$ to values of order unity~\cite{Yueetal12}.\\
%
%
In Fig.~\ref{fig:2}-a we illustrate the effects introduced by the polymers for the aforementioned range of $\tau_P$ for a fixed $L^2=5 \times 10^3$: at low $\mathrm{Ca}$, the linear behaviour of the $\mathrm{Ca}$ \textit{vs.} $\mathrm{Bo}$ curve is recovered for all the data, which nicely collapse on the same master curve. By increasing the driving force, however, a critical $\mathrm{Ca}$ exists around which the linear behaviour that is found for a Newtonian fluid is spoiled by viscoelasticity and a plateau emerges. These results are complemented in Fig.~\ref{fig:2}-b that shows the effect of an increase in $L^2$ for a fixed $\tau_P$. It is readily verified that the deviations from linearity first emerge with a slight sublinear behaviour for the smallest $L^2$, while they produce a plateau for the largest $L^2$ ($L^2 \approx \mathcal{O}(10^3)$) \cite{Lindner03,Arratia}. Overall, it is important to observe that the characteristic Ca at which the linear behaviour starts to be violated is only weakly dependent on the value of the finite extensibility parameter, while it is more sensitive to the value of the polymer relaxation time. 
The characteristic velocity $U_c$ at which the linear behaviour starts to be violated can be deduced from these data. We estimated that the linear behaviour is indeed lost when $\tau_P/\tau_{\tiny\mbox{fluid}} \approx 1$, i.e. when the ratio of the polymer relaxation time to the characteristic time of the fluid in the wedge close to the contact line is of the order one.\\ 
With respect to the experimental findings of Fig.~\ref{fig:exp_PAA_glycerol} few important remarks are in order. Numerical results of Fig.~\ref{fig:2}-a show that the larger the polymer relaxation time and the smaller is the velocity at which the sublinear behaviour emerges. Considering the rheological data of Fig.~\ref{fig:visco}, and using a fitting procedure on normal stresses~\cite{Arratia}, one could estimate that the Boger fluid and PAA$_{\textrm{HM}}$/water solution 2500 ppm exhibit a larger relaxation time in comparison to the PAA$_{\textrm{LM}}$/water solution 10000 ppm. Then, based on the results of Fig.~\ref{fig:2}-a, deviations from linearity should start at smaller velocities. This is not what observed in experiments. Following other studies in the literature~\cite{Amberg,Ladd05,GraamReview}, one can speculate that wall boundary effects~\cite{MaGraam05,Ladd05,GraamReview} play a role. Polymers under the effect of shear may migrate away from the wedge, with the migration more pronounced at increasing elasticity. This would have no counterpart in a model description based on conformation tensor~\cite{Amberg}, like the one we used. 
In addition, numerical simulations, by construction,  are not affected by hysteresis effects (i.e. $Bo_{c}=0$)~\cite{PRLnoi,PREnoi,Langmuir,EPJnoi}, whereas experiments experience them. 
It is then possible that the pinning-depinning point has some role in ``triggering'' the drop elongation. We hasten to remark that these are open points worth to be elucidated with future studies.\\
Let us then discuss the shape of the drops during the sliding and their relation to the measured velocity. The slight sublinear behaviour observed for $L^2=10^2$ in Fig.~\ref{fig:2}-b is an indication that the driving force (gravity) is balanced by two distinct effects, one linear in the velocity (viscous dissipation) and a non-linear one. However, in such conditions, we did not observe a remarkable change in the drop shape. This contrasts the situation at larger $L^2$ where we observed the plateau behaviour, which somehow signals a ``transition'' in the dynamics of the sliding drop. This is evident in Fig.~\ref{fig:3}, where we report a simultaneous view of the stationary drop shapes together with snapshots of the polymer feedback stresses for a fixed $L^2= 5 \times 10^3$ and fixed $\mathrm{{Bo}=0.024}$, by varying the polymer relaxation time $\tau_P$. We observed pronounced viscoelastic effects in the wedge flow close to the contact line while the bulk of the drop behaved essentially as a Newtonian fluid. Normal stresses basically introduced an extra driving force which caused drop elongation in correspondence of the point where we observed a flattening of the Ca {\it vs.} Bo curve.\\
Results of Fig.~\ref{fig:3} are finally complemented with the data reported in Fig.~\ref{fig:4}, where we show the time history of the drop length $L_d$ (computed as the distance between the front and rear contact line) normalised to the drop diameter at rest for Newtonian and non-Newtonian cases and different Bo numbers. For all the Newtonian cases, the stationary drop length did not vary. Conversely, for the non-Newtonian cases above a given Bo number (corresponding to the critical Ca number), the drop length started to increase and reached a stationary state over a time lapse of a few tens of $\tau_P$ for the largest Bo numbers considered.

\section{Conclusions} \label{conclusion}

In this paper we investigated non-Newtonian, viscoelastic drops made of flexible polymers sliding down a homogeneous inclined surface. In our study we considered the effects of either the shear-thinning or the elasticity of the polymer solutions, and we also reviewed previous results obtained with stiff polymers~\cite{EPJnoi}. 
We first analysed the relation between the Capillary number and Bond number and provided quantitative details on how such relation is affected by different definitions of the Ca number. Drops with flexible polymers exhibit a remarkable stretching in steady sliding, which strongly contrasts with what is observed in drops with stiff polymers, where the elongation is not observed even at the highest concentrations. This may be attributed to the interface bending effects induced by viscosity \cite{snoeijer_moving_2013} which are reduced by the strong shear-thinning \cite{WeiGaroff07,WeiGaroff09}, although a more quantitative assessment of this finding requires future investigations. 
On the other hand, for flexible polymers, we found clear evidences that drop elongation requires a sort of ``optimal elasticity'' \cite{WeiGaroff07,WeiGaroff09,Yueetal12} to be observed.\\
To complement the experimental data and relate the experimental observations to the micromechanics of the polymers, we used numerical simulations of an ``idealised'' drop with a dilute polymeric solution of non interacting Finitely Extensible Nonlinear Elastic dumbbells (FENE-P model). Specifically, we extended the numerical analysis of \cite{EPJnoi} by performing a study in terms of the different model parameters (mainly the polymer relaxation time $\tau_P$ and their maximum squared extensibility $L^2$) to highlight how the emergence of the observed experimental behaviours could be related to these micromechanical details. We could also visualise the distribution of the polymer feedback stresses during the motion of the drop, thus correlating the distribution of those stresses to the interface shape and the resulting macroscopic velocity.\\


\begin{acknowledgments}
The authors kindly acknowledge funding from the European Research Council under the European Community's Seventh Framework Programme (FP7/2007-2013)/ERC Grant Agreement No. 279004 (DROEMU). MS acknowledges the European Union's Framework Programme for Research and Innovation Horizon 2020 (2014-2020) under the Marie Sklodowska-Curie Grant Agreement No. 642069 (HPC-LEAP). We are particularly grateful to Michele Minchio, Chiara Guidolin and Rasa Platakyte for their help in the experimental work. In addition, we are indebted to Prof. Alessandro Martucci (University of Padova) and Prof. Giovanni Lucchetta (Department of Industrial Engineering, University of Padova) for the fruitful discussions and support about rheological characterisation. We acknowledge Anupam Gupta for support in the analysis of the numerical simulations. MS acknowledges useful discussions and fruitful exchange of ideas with Ciro Semprebon and Martin Brinkmann.
\end{acknowledgments}

\providecommand*{\mcitethebibliography}{\thebibliography}
\csname @ifundefined\endcsname{endmcitethebibliography}
{\let\endmcitethebibliography\endthebibliography}{}



\begin{mcitethebibliography}{70}
\providecommand*{\natexlab}[1]{#1}
\providecommand*{\mciteSetBstSublistMode}[1]{}
\providecommand*{\mciteSetBstMaxWidthForm}[2]{}
\providecommand*{\mciteBstWouldAddEndPuncttrue}
  {\def\EndOfBibitem{\unskip.}}
\providecommand*{\mciteBstWouldAddEndPunctfalse}
  {\let\EndOfBibitem\relax}
\providecommand*{\mciteSetBstMidEndSepPunct}[3]{}
\providecommand*{\mciteSetBstSublistLabelBeginEnd}[3]{}
\providecommand*{\EndOfBibitem}{}
\mciteSetBstSublistMode{f}
\mciteSetBstMaxWidthForm{subitem}
{(\emph{\alph{mcitesubitemcount}})}
\mciteSetBstSublistLabelBeginEnd{\mcitemaxwidthsubitemform\space}
{\relax}{\relax}

\bibitem[Qu{\'e}r{\'e}(2008)]{quere08}
D.~Qu{\'e}r{\'e}, \emph{Annu. Rev. Mater. Res.}, 2008, \textbf{38},
  71--99\relax
\mciteBstWouldAddEndPuncttrue
\mciteSetBstMidEndSepPunct{\mcitedefaultmidpunct}
{\mcitedefaultendpunct}{\mcitedefaultseppunct}\relax
\EndOfBibitem
\bibitem[Bonn \emph{et~al.}(2009)Bonn, Eggers, Indekeu, Meunier, and
  Rolley]{Bonn09}
D.~Bonn, J.~Eggers, J.~Indekeu, J.~Meunier and E.~Rolley, \emph{Rev. Mod.
  Phys.}, 2009, \textbf{81}, 739\relax
\mciteBstWouldAddEndPuncttrue
\mciteSetBstMidEndSepPunct{\mcitedefaultmidpunct}
{\mcitedefaultendpunct}{\mcitedefaultseppunct}\relax
\EndOfBibitem
\bibitem[Liu \emph{et~al.}(2010)Liu, Yao, and Jiang]{Liu10}
K.~Liu, X.~Yao and L.~Jiang, \emph{Chem. Soc. Rev.}, 2010, \textbf{39},
  3240--3255\relax
\mciteBstWouldAddEndPuncttrue
\mciteSetBstMidEndSepPunct{\mcitedefaultmidpunct}
{\mcitedefaultendpunct}{\mcitedefaultseppunct}\relax
\EndOfBibitem
\bibitem[Bussonni\`ere \emph{et~al.}(2016)Bussonni\`ere, Baudoin, Brunet, and
  Matar]{Brunet16}
A.~Bussonni\`ere, M.~Baudoin, P.~Brunet and O.~B. Matar, \emph{Phys. Rev. E},
  2016, \textbf{93}, 053106\relax
\mciteBstWouldAddEndPuncttrue
\mciteSetBstMidEndSepPunct{\mcitedefaultmidpunct}
{\mcitedefaultendpunct}{\mcitedefaultseppunct}\relax
\EndOfBibitem
\bibitem[Sakai \emph{et~al.}(2016)Sakai, Kato, Ishizuka, Isobe, Nakajima, and
  Fujishima]{Nakajima16}
M.~Sakai, T.~Kato, N.~Ishizuka, T.~Isobe, A.~Nakajima and A.~Fujishima,
  \emph{J. Sol-Gel Sci. Techn.}, 2016, \textbf{77}, 257--265\relax
\mciteBstWouldAddEndPuncttrue
\mciteSetBstMidEndSepPunct{\mcitedefaultmidpunct}
{\mcitedefaultendpunct}{\mcitedefaultseppunct}\relax
\EndOfBibitem
\bibitem[Semprebon \emph{et~al.}(2016)Semprebon, Varagnolo, Filippi, Perlini,
  Pierno, Brinkmann, and Mistura]{Semprebon16}
C.~Semprebon, S.~Varagnolo, D.~Filippi, L.~Perlini, M.~Pierno, M.~Brinkmann and
  G.~Mistura, \emph{Soft Matter}, 2016, \textbf{12}, 8268--8273\relax
\mciteBstWouldAddEndPuncttrue
\mciteSetBstMidEndSepPunct{\mcitedefaultmidpunct}
{\mcitedefaultendpunct}{\mcitedefaultseppunct}\relax
\EndOfBibitem
\bibitem[Barbulovic-Nad \emph{et~al.}(2010)Barbulovic-Nad, Au, and
  Wheeler]{wheeler10}
I.~Barbulovic-Nad, S.~H. Au and A.~R. Wheeler, \emph{Lab. Chip}, 2010,
  \textbf{10}, 1536--1542\relax
\mciteBstWouldAddEndPuncttrue
\mciteSetBstMidEndSepPunct{\mcitedefaultmidpunct}
{\mcitedefaultendpunct}{\mcitedefaultseppunct}\relax
\EndOfBibitem
\bibitem[Mousa \emph{et~al.}(2009)Mousa, Jebrail, Yang, Abdelgawad, Metalnikov,
  Chen, Wheeler, and Casper]{mousa09}
N.~A. Mousa, M.~J. Jebrail, H.~Yang, M.~Abdelgawad, P.~Metalnikov, J.~Chen,
  A.~R. Wheeler and R.~F. Casper, \emph{Science Translational Medicine}, 2009,
  \textbf{1}, 1ra2--1ra2\relax
\mciteBstWouldAddEndPuncttrue
\mciteSetBstMidEndSepPunct{\mcitedefaultmidpunct}
{\mcitedefaultendpunct}{\mcitedefaultseppunct}\relax
\EndOfBibitem
\bibitem[Laan \emph{et~al.}(2014)Laan, de~Bruin, Bartolo, Josserand, and
  Bonn]{laan14}
N.~Laan, K.~G. de~Bruin, D.~Bartolo, C.~Josserand and D.~Bonn, \emph{Phy. Rev.
  Applied}, 2014, \textbf{2}, 044018\relax
\mciteBstWouldAddEndPuncttrue
\mciteSetBstMidEndSepPunct{\mcitedefaultmidpunct}
{\mcitedefaultendpunct}{\mcitedefaultseppunct}\relax
\EndOfBibitem
\bibitem[Shih \emph{et~al.}(2012)Shih, Yang, Jebrail, Fobel, McIntosh,
  Al-Dirbashi, Chakraborty, and Wheeler]{wheeler12}
S.~C. Shih, H.~Yang, M.~J. Jebrail, R.~Fobel, N.~McIntosh, O.~Y. Al-Dirbashi,
  P.~Chakraborty and A.~R. Wheeler, \emph{Anal. Chem.}, 2012, \textbf{84},
  3731--3738\relax
\mciteBstWouldAddEndPuncttrue
\mciteSetBstMidEndSepPunct{\mcitedefaultmidpunct}
{\mcitedefaultendpunct}{\mcitedefaultseppunct}\relax
\EndOfBibitem
\bibitem[Son and Kim(2009)]{son09}
Y.~Son and C.~Kim, \emph{J. Non-Newton. Fluid.}, 2009, \textbf{162},
  78--87\relax
\mciteBstWouldAddEndPuncttrue
\mciteSetBstMidEndSepPunct{\mcitedefaultmidpunct}
{\mcitedefaultendpunct}{\mcitedefaultseppunct}\relax
\EndOfBibitem
\bibitem[Rafai and Bonn(2005)]{Rafai05}
S.~Rafai and D.~Bonn, \emph{Physica A}, 2005, \textbf{358}, 58--67\relax
\mciteBstWouldAddEndPuncttrue
\mciteSetBstMidEndSepPunct{\mcitedefaultmidpunct}
{\mcitedefaultendpunct}{\mcitedefaultseppunct}\relax
\EndOfBibitem
\bibitem[Rafai \emph{et~al.}(2004)Rafai, Bonn, and Boudaoud]{Rafai}
S.~Rafai, D.~Bonn and A.~Boudaoud, \emph{J. Fluid Mech.}, 2004, \textbf{513},
  77--85\relax
\mciteBstWouldAddEndPuncttrue
\mciteSetBstMidEndSepPunct{\mcitedefaultmidpunct}
{\mcitedefaultendpunct}{\mcitedefaultseppunct}\relax
\EndOfBibitem
\bibitem[Carr\'{e} and Eustache(1997)]{Carre97}
A.~Carr\'{e} and F.~Eustache, \emph{C. R. Acad. Sci. Paris}, 1997,
  \textbf{325}, 709--718\relax
\mciteBstWouldAddEndPuncttrue
\mciteSetBstMidEndSepPunct{\mcitedefaultmidpunct}
{\mcitedefaultendpunct}{\mcitedefaultseppunct}\relax
\EndOfBibitem
\bibitem[Han and Kim(2013)]{HanKim13}
J.~Han and C.~Kim, \emph{J. Non-Newton. Fluid.}, 2013, \textbf{202},
  120--130\relax
\mciteBstWouldAddEndPuncttrue
\mciteSetBstMidEndSepPunct{\mcitedefaultmidpunct}
{\mcitedefaultendpunct}{\mcitedefaultseppunct}\relax
\EndOfBibitem
\bibitem[Min \emph{et~al.}(2010)Min, Duan, Wang, Liang, Lee, and Su]{min10}
Q.~Min, Y.-Y. Duan, X.-D. Wang, Z.-P. Liang, D.-J. Lee and A.~Su, \emph{J.
  Colloid Interf. Sci.}, 2010, \textbf{348}, 250--254\relax
\mciteBstWouldAddEndPuncttrue
\mciteSetBstMidEndSepPunct{\mcitedefaultmidpunct}
{\mcitedefaultendpunct}{\mcitedefaultseppunct}\relax
\EndOfBibitem
\bibitem[Wei \emph{et~al.}(2009)Wei, Rame, Walker, and Garoff]{wei09}
Y.~Wei, E.~Rame, L.~Walker and S.~Garoff, \emph{J. Phys-Condens. Mat.}, 2009,
  \textbf{21}, 464126\relax
\mciteBstWouldAddEndPuncttrue
\mciteSetBstMidEndSepPunct{\mcitedefaultmidpunct}
{\mcitedefaultendpunct}{\mcitedefaultseppunct}\relax
\EndOfBibitem
\bibitem[Wei \emph{et~al.}(2007)Wei, Seevaratnam, Garoff, Ram{\'e}, and
  Walker]{wei07}
Y.~Wei, G.~Seevaratnam, S.~Garoff, E.~Ram{\'e} and L.~Walker, \emph{J. Colloid
  Interf. Sci.}, 2007, \textbf{313}, 274--280\relax
\mciteBstWouldAddEndPuncttrue
\mciteSetBstMidEndSepPunct{\mcitedefaultmidpunct}
{\mcitedefaultendpunct}{\mcitedefaultseppunct}\relax
\EndOfBibitem
\bibitem[Seevaratnam \emph{et~al.}(2007)Seevaratnam, Suo, Ram{\'e}, Walker, and
  Garoff]{wei07thinning}
G.~Seevaratnam, Y.~Suo, E.~Ram{\'e}, L.~Walker and S.~Garoff, \emph{Phys.
  Fluids}, 2007, \textbf{19}, 012103\relax
\mciteBstWouldAddEndPuncttrue
\mciteSetBstMidEndSepPunct{\mcitedefaultmidpunct}
{\mcitedefaultendpunct}{\mcitedefaultseppunct}\relax
\EndOfBibitem
\bibitem[Wang \emph{et~al.}(2007)Wang, Lee, Peng, and Lai]{wang07}
X.~Wang, D.~Lee, X.~Peng and J.~Lai, \emph{Langmuir}, 2007, \textbf{23},
  8042--8047\relax
\mciteBstWouldAddEndPuncttrue
\mciteSetBstMidEndSepPunct{\mcitedefaultmidpunct}
{\mcitedefaultendpunct}{\mcitedefaultseppunct}\relax
\EndOfBibitem
\bibitem[Boudaoud(2007)]{boudaoud07}
A.~Boudaoud, \emph{Eur. Phys. J. E}, 2007, \textbf{22}, 107--109\relax
\mciteBstWouldAddEndPuncttrue
\mciteSetBstMidEndSepPunct{\mcitedefaultmidpunct}
{\mcitedefaultendpunct}{\mcitedefaultseppunct}\relax
\EndOfBibitem
\bibitem[Han and Kim(2014)]{HanKim14}
J.~Han and C.~Kim, \emph{Rheol. Acta.}, 2014, \textbf{53}, 55--66\relax
\mciteBstWouldAddEndPuncttrue
\mciteSetBstMidEndSepPunct{\mcitedefaultmidpunct}
{\mcitedefaultendpunct}{\mcitedefaultseppunct}\relax
\EndOfBibitem
\bibitem[Liu \emph{et~al.}(2009)Liu, Bonaccurso, Sokuler, Auernhammer, and
  Butt]{liu09}
C.~Liu, E.~Bonaccurso, M.~Sokuler, G.~K. Auernhammer and H.-J. Butt,
  \emph{Langmuir}, 2009, \textbf{26}, 2544--2549\relax
\mciteBstWouldAddEndPuncttrue
\mciteSetBstMidEndSepPunct{\mcitedefaultmidpunct}
{\mcitedefaultendpunct}{\mcitedefaultseppunct}\relax
\EndOfBibitem
\bibitem[Min \emph{et~al.}(2013)Min, Duan, Wang, Liang, and Lee]{min13}
Q.~Min, Y.-Y. Duan, X.-D. Wang, Z.-P. Liang and D.-J. Lee, \emph{Int. J.
  Thermophys.}, 2013, \textbf{34}, 2276--2285\relax
\mciteBstWouldAddEndPuncttrue
\mciteSetBstMidEndSepPunct{\mcitedefaultmidpunct}
{\mcitedefaultendpunct}{\mcitedefaultseppunct}\relax
\EndOfBibitem
\bibitem[Nakajima(2011)]{Nakajima11}
A.~Nakajima, \emph{NPG Asia Mater.}, 2011, \textbf{3}, 49--56\relax
\mciteBstWouldAddEndPuncttrue
\mciteSetBstMidEndSepPunct{\mcitedefaultmidpunct}
{\mcitedefaultendpunct}{\mcitedefaultseppunct}\relax
\EndOfBibitem
\bibitem[Boyer \emph{et~al.}(2016)Boyer, Sandoval-Nava, Snoeijer, Dijksman, and
  Lohse]{boyer16}
F.~Boyer, E.~Sandoval-Nava, J.~H. Snoeijer, J.~F. Dijksman and D.~Lohse,
  \emph{Phys. Rev. Fluids}, 2016, \textbf{1}, 013901\relax
\mciteBstWouldAddEndPuncttrue
\mciteSetBstMidEndSepPunct{\mcitedefaultmidpunct}
{\mcitedefaultendpunct}{\mcitedefaultseppunct}\relax
\EndOfBibitem
\bibitem[Zang \emph{et~al.}(2014)Zang, Zhang, Song, Chen, Zhang, Geng, and
  Chen]{zang14}
D.~Zang, W.~Zhang, J.~Song, Z.~Chen, Y.~Zhang, X.~Geng and F.~Chen, \emph{Appl.
  Phys. Lett.}, 2014, \textbf{105}, 231603\relax
\mciteBstWouldAddEndPuncttrue
\mciteSetBstMidEndSepPunct{\mcitedefaultmidpunct}
{\mcitedefaultendpunct}{\mcitedefaultseppunct}\relax
\EndOfBibitem
\bibitem[Zang \emph{et~al.}(2013)Zang, Wang, Geng, Zhang, and Chen]{zang13}
D.~Zang, X.~Wang, X.~Geng, Y.~Zhang and Y.~Chen, \emph{Soft Matter}, 2013,
  \textbf{9}, 394--400\relax
\mciteBstWouldAddEndPuncttrue
\mciteSetBstMidEndSepPunct{\mcitedefaultmidpunct}
{\mcitedefaultendpunct}{\mcitedefaultseppunct}\relax
\EndOfBibitem
\bibitem[Ravi \emph{et~al.}(2013)Ravi, Jog, and Manglik]{ravi13}
V.~Ravi, M.~A. Jog and R.~M. Manglik, \emph{Interfacial Phenomena and Heat
  Transfer}, 2013, \textbf{1}, year\relax
\mciteBstWouldAddEndPuncttrue
\mciteSetBstMidEndSepPunct{\mcitedefaultmidpunct}
{\mcitedefaultendpunct}{\mcitedefaultseppunct}\relax
\EndOfBibitem
\bibitem[An and Lee(2012)]{an12}
S.~M. An and S.~Y. Lee, \emph{Exp. Therm. Fluid Sci.}, 2012, \textbf{37},
  37--45\relax
\mciteBstWouldAddEndPuncttrue
\mciteSetBstMidEndSepPunct{\mcitedefaultmidpunct}
{\mcitedefaultendpunct}{\mcitedefaultseppunct}\relax
\EndOfBibitem
\bibitem[Huh \emph{et~al.}(2015)Huh, Jung, Seo, and Lee]{huh15}
H.~K. Huh, S.~Jung, K.~W. Seo and S.~J. Lee, \emph{Microfluid. Nanofluid.},
  2015, \textbf{18}, 1221--1232\relax
\mciteBstWouldAddEndPuncttrue
\mciteSetBstMidEndSepPunct{\mcitedefaultmidpunct}
{\mcitedefaultendpunct}{\mcitedefaultseppunct}\relax
\EndOfBibitem
\bibitem[Varagnolo \emph{et~al.}(2015)Varagnolo, Mistura, Pierno, and
  Sbragaglia]{EPJnoi}
S.~Varagnolo, G.~Mistura, M.~Pierno and M.~Sbragaglia, \emph{Eur. Phys. J. E},
  2015, \textbf{38}, 1--8\relax
\mciteBstWouldAddEndPuncttrue
\mciteSetBstMidEndSepPunct{\mcitedefaultmidpunct}
{\mcitedefaultendpunct}{\mcitedefaultseppunct}\relax
\EndOfBibitem
\bibitem[Morita \emph{et~al.}(2009)Morita, Plog, Kajiya, and M.]{Morita09}
H.~Morita, S.~Plog, T.~Kajiya and D.~M., \emph{J. Phys. Soc. Jpn.}, 2009,
  \textbf{78}, 014804\relax
\mciteBstWouldAddEndPuncttrue
\mciteSetBstMidEndSepPunct{\mcitedefaultmidpunct}
{\mcitedefaultendpunct}{\mcitedefaultseppunct}\relax
\EndOfBibitem
\bibitem[Seevaratnam \emph{et~al.}(2007)Seevaratnam, Suo, Ram\'{e}, Walker, and
  Garoff]{Seevaratnam07}
G.~Seevaratnam, Y.~Suo, E.~Ram\'{e}, L.~Walker and S.~Garoff, \emph{Phys.
  Fluids}, 2007, \textbf{19}, 012103\relax
\mciteBstWouldAddEndPuncttrue
\mciteSetBstMidEndSepPunct{\mcitedefaultmidpunct}
{\mcitedefaultendpunct}{\mcitedefaultseppunct}\relax
\EndOfBibitem
\bibitem[German and Bertola(2010)]{german10}
G.~German and V.~Bertola, \emph{Colloid. Surface. A}, 2010, \textbf{366},
  18--26\relax
\mciteBstWouldAddEndPuncttrue
\mciteSetBstMidEndSepPunct{\mcitedefaultmidpunct}
{\mcitedefaultendpunct}{\mcitedefaultseppunct}\relax
\EndOfBibitem
\bibitem[Moon \emph{et~al.}(2014)Moon, Kim, and Lee]{moon14}
J.~H. Moon, D.~Y. Kim and S.~H. Lee, \emph{Exp. Therm. Fluid Sci.}, 2014,
  \textbf{57}, 94--101\relax
\mciteBstWouldAddEndPuncttrue
\mciteSetBstMidEndSepPunct{\mcitedefaultmidpunct}
{\mcitedefaultendpunct}{\mcitedefaultseppunct}\relax
\EndOfBibitem
\bibitem[Smith and Bertola(2010)]{smith10}
M.~Smith and V.~Bertola, \emph{Phys. Rev. Lett.}, 2010, \textbf{104},
  154502\relax
\mciteBstWouldAddEndPuncttrue
\mciteSetBstMidEndSepPunct{\mcitedefaultmidpunct}
{\mcitedefaultendpunct}{\mcitedefaultseppunct}\relax
\EndOfBibitem
\bibitem[Smith and Sharp(2014)]{smith14}
M.~Smith and J.~Sharp, \emph{Langmuir}, 2014, \textbf{30}, 5455--5459\relax
\mciteBstWouldAddEndPuncttrue
\mciteSetBstMidEndSepPunct{\mcitedefaultmidpunct}
{\mcitedefaultendpunct}{\mcitedefaultseppunct}\relax
\EndOfBibitem
\bibitem[Whitcomb and Macosko(1978)]{Macosko78}
P.~J. Whitcomb and C.~W. Macosko, \emph{J. Rheol}, 1978, \textbf{22}, 493\relax
\mciteBstWouldAddEndPuncttrue
\mciteSetBstMidEndSepPunct{\mcitedefaultmidpunct}
{\mcitedefaultendpunct}{\mcitedefaultseppunct}\relax
\EndOfBibitem
\bibitem[Callaghan and Gil(2000)]{Callaghan00}
P.~T. Callaghan and A.~M. Gil, \emph{Macromolecules}, 2000, \textbf{33},
  4116--4124\relax
\mciteBstWouldAddEndPuncttrue
\mciteSetBstMidEndSepPunct{\mcitedefaultmidpunct}
{\mcitedefaultendpunct}{\mcitedefaultseppunct}\relax
\EndOfBibitem
\bibitem[Tanner(2000)]{tanner}
R.~I. Tanner, \emph{Engineering rheology}, OUP Oxford, 2000, vol.~52\relax
\mciteBstWouldAddEndPuncttrue
\mciteSetBstMidEndSepPunct{\mcitedefaultmidpunct}
{\mcitedefaultendpunct}{\mcitedefaultseppunct}\relax
\EndOfBibitem
\bibitem[Arratia \emph{et~al.}(2009)Arratia, Cramer, Gollub, and
  Durian]{Arratia}
P.~Arratia, L.-A. Cramer, J.~Gollub and D.~J. Durian, \emph{New J. Phys.},
  2009, \textbf{11}, 115006\relax
\mciteBstWouldAddEndPuncttrue
\mciteSetBstMidEndSepPunct{\mcitedefaultmidpunct}
{\mcitedefaultendpunct}{\mcitedefaultseppunct}\relax
\EndOfBibitem
\bibitem[Derzsi \emph{et~al.}(2013)Derzsi, Kasprzyk, Plog, and
  Garstecki]{derzsi2013flow}
L.~Derzsi, M.~Kasprzyk, J.~P. Plog and P.~Garstecki, \emph{Phys. Fluids}, 2013,
  \textbf{25}, 092001\relax
\mciteBstWouldAddEndPuncttrue
\mciteSetBstMidEndSepPunct{\mcitedefaultmidpunct}
{\mcitedefaultendpunct}{\mcitedefaultseppunct}\relax
\EndOfBibitem
\bibitem[Toth \emph{et~al.}(2011)Toth, Ferraro, Chiarello, Pierno, Mistura,
  Bissacco, and Semprebon]{Toth11}
T.~Toth, D.~Ferraro, E.~Chiarello, M.~Pierno, G.~Mistura, G.~Bissacco and
  C.~Semprebon, \emph{Langmuir}, 2011, \textbf{27}, 4742--4748\relax
\mciteBstWouldAddEndPuncttrue
\mciteSetBstMidEndSepPunct{\mcitedefaultmidpunct}
{\mcitedefaultendpunct}{\mcitedefaultseppunct}\relax
\EndOfBibitem
\bibitem[Varagnolo \emph{et~al.}(2013)Varagnolo, Ferraro, Fantinel, Pierno,
  Mistura, Amati, Biferale, and Sbragaglia]{PRLnoi}
S.~Varagnolo, D.~Ferraro, P.~Fantinel, M.~Pierno, G.~Mistura, G.~Amati,
  L.~Biferale and M.~Sbragaglia, \emph{Phys. Rev. Lett.}, 2013, \textbf{111},
  066101\relax
\mciteBstWouldAddEndPuncttrue
\mciteSetBstMidEndSepPunct{\mcitedefaultmidpunct}
{\mcitedefaultendpunct}{\mcitedefaultseppunct}\relax
\EndOfBibitem
\bibitem[Sbragaglia \emph{et~al.}(2014)Sbragaglia, Biferale, Amati, Varagnolo,
  Ferraro, Mistura, and Pierno]{PREnoi}
M.~Sbragaglia, L.~Biferale, G.~Amati, S.~Varagnolo, D.~Ferraro, G.~Mistura and
  M.~Pierno, \emph{Phys. Rev. E}, 2014, \textbf{89}, 012406\relax
\mciteBstWouldAddEndPuncttrue
\mciteSetBstMidEndSepPunct{\mcitedefaultmidpunct}
{\mcitedefaultendpunct}{\mcitedefaultseppunct}\relax
\EndOfBibitem
\bibitem[Varagnolo \emph{et~al.}(2014)Varagnolo, Schiocchet, Ferraro, Pierno,
  Mistura, Sbragaglia, Gupta, and Amati]{Langmuir}
S.~Varagnolo, V.~Schiocchet, D.~Ferraro, M.~Pierno, G.~Mistura, M.~Sbragaglia,
  A.~Gupta and G.~Amati, \emph{Langmuir}, 2014, \textbf{30}, 2401--2409\relax
\mciteBstWouldAddEndPuncttrue
\mciteSetBstMidEndSepPunct{\mcitedefaultmidpunct}
{\mcitedefaultendpunct}{\mcitedefaultseppunct}\relax
\EndOfBibitem
\bibitem[Podgorski \emph{et~al.}(2001)Podgorski, Flesselles, and
  Limat]{Podgorski01}
T.~Podgorski, J.-M. Flesselles and L.~Limat, \emph{Phys. Rev. Lett.}, 2001,
  \textbf{87}, 036102\relax
\mciteBstWouldAddEndPuncttrue
\mciteSetBstMidEndSepPunct{\mcitedefaultmidpunct}
{\mcitedefaultendpunct}{\mcitedefaultseppunct}\relax
\EndOfBibitem
\bibitem[Kim \emph{et~al.}(2002)Kim, Lee, and Kang]{Kimetal02}
H.~Kim, H.~Lee and B.~Kang, \emph{J. Colloid Interf. Sci.}, 2002, \textbf{247},
  372--380\relax
\mciteBstWouldAddEndPuncttrue
\mciteSetBstMidEndSepPunct{\mcitedefaultmidpunct}
{\mcitedefaultendpunct}{\mcitedefaultseppunct}\relax
\EndOfBibitem
\bibitem[Le~Grand \emph{et~al.}(2005)Le~Grand, Daerr, and Limat]{Legrand05}
N.~Le~Grand, A.~Daerr and L.~Limat, \emph{J. Fluid Mech.}, 2005, \textbf{541},
  293--315\relax
\mciteBstWouldAddEndPuncttrue
\mciteSetBstMidEndSepPunct{\mcitedefaultmidpunct}
{\mcitedefaultendpunct}{\mcitedefaultseppunct}\relax
\EndOfBibitem
\bibitem[Furmidge(1962)]{Furmidge62}
C.~Furmidge, \emph{J. Colloid Interf. Sci.}, 1962, \textbf{17}, 309--324\relax
\mciteBstWouldAddEndPuncttrue
\mciteSetBstMidEndSepPunct{\mcitedefaultmidpunct}
{\mcitedefaultendpunct}{\mcitedefaultseppunct}\relax
\EndOfBibitem
\bibitem[Bird \emph{et~al.}(1980)Bird, Dotson, and Armstrong]{birdpaper}
R.~Bird, P.~Dotson and R.~Armstrong, \emph{J. Non-Newton. Fluid.}, 1980,
  \textbf{7}, 213--235\relax
\mciteBstWouldAddEndPuncttrue
\mciteSetBstMidEndSepPunct{\mcitedefaultmidpunct}
{\mcitedefaultendpunct}{\mcitedefaultseppunct}\relax
\EndOfBibitem
\bibitem[Bird \emph{et~al.}(1987)Bird, Armstrong, and Hassager]{bird87}
R.~B. Bird, R.~C. Armstrong and O.~Hassager, \emph{Dynamics of polymeric
  liquids}, J. Wiley \& Sons, 1987\relax
\mciteBstWouldAddEndPuncttrue
\mciteSetBstMidEndSepPunct{\mcitedefaultmidpunct}
{\mcitedefaultendpunct}{\mcitedefaultseppunct}\relax
\EndOfBibitem
\bibitem[Herrchen and Oettinger(1997)]{Herrchen97}
M.~Herrchen and H.~Oettinger, \emph{J. Non-Newton. Fluid.}, 1997, \textbf{68},
  17--42\relax
\mciteBstWouldAddEndPuncttrue
\mciteSetBstMidEndSepPunct{\mcitedefaultmidpunct}
{\mcitedefaultendpunct}{\mcitedefaultseppunct}\relax
\EndOfBibitem
\bibitem[Wagner \emph{et~al.}(2005)Wagner, Amarouchene, Bonn, and
  Eggers]{Wagner05}
C.~Wagner, Y.~Amarouchene, D.~Bonn and J.~Eggers, \emph{Phys. Rev. Lett.},
  2005, \textbf{95}, 164504\relax
\mciteBstWouldAddEndPuncttrue
\mciteSetBstMidEndSepPunct{\mcitedefaultmidpunct}
{\mcitedefaultendpunct}{\mcitedefaultseppunct}\relax
\EndOfBibitem
\bibitem[Lindner \emph{et~al.}(2003)Lindner, Vermant, and Bonn]{Lindner03}
A.~Lindner, J.~Vermant and D.~Bonn, \emph{Physica A}, 2003, \textbf{319},
  125--133\relax
\mciteBstWouldAddEndPuncttrue
\mciteSetBstMidEndSepPunct{\mcitedefaultmidpunct}
{\mcitedefaultendpunct}{\mcitedefaultseppunct}\relax
\EndOfBibitem
\bibitem[Moradi \emph{et~al.}(2011)Moradi, Varnik, and Steinbach]{Moradi}
N.~Moradi, F.~Varnik and I.~Steinbach, \emph{Europhys. Lett.}, 2011,
  \textbf{95}, 44003\relax
\mciteBstWouldAddEndPuncttrue
\mciteSetBstMidEndSepPunct{\mcitedefaultmidpunct}
{\mcitedefaultendpunct}{\mcitedefaultseppunct}\relax
\EndOfBibitem
\bibitem[Kusumaatmaja \emph{et~al.}(2006)Kusumaatmaja, Leopoldes, Dupuis, and
  Yeomans]{Kusumaatmaja}
H.~Kusumaatmaja, J.~Leopoldes, A.~Dupuis and J.~Yeomans, \emph{Europhys.
  Lett.}, 2006, \textbf{73}, 740\relax
\mciteBstWouldAddEndPuncttrue
\mciteSetBstMidEndSepPunct{\mcitedefaultmidpunct}
{\mcitedefaultendpunct}{\mcitedefaultseppunct}\relax
\EndOfBibitem
\bibitem[Gupta \emph{et~al.}(2015)Gupta, Sbragaglia, and
  Scagliarini]{SbragagliaGuptaScagliarini}
A.~Gupta, M.~Sbragaglia and A.~Scagliarini, \emph{J. Comput. Phys.}, 2015,
  \textbf{291}, 177--197\relax
\mciteBstWouldAddEndPuncttrue
\mciteSetBstMidEndSepPunct{\mcitedefaultmidpunct}
{\mcitedefaultendpunct}{\mcitedefaultseppunct}\relax
\EndOfBibitem
\bibitem[Gupta and Sbragaglia(2014)]{SbragagliaGupta}
A.~Gupta and M.~Sbragaglia, \emph{Phys. Rev. E}, 2014, \textbf{90},
  023305\relax
\mciteBstWouldAddEndPuncttrue
\mciteSetBstMidEndSepPunct{\mcitedefaultmidpunct}
{\mcitedefaultendpunct}{\mcitedefaultseppunct}\relax
\EndOfBibitem
\bibitem[Seevaratnam \emph{et~al.}(2005)Seevaratnam, Walker, Ram\'{e}, and
  Garoff]{Seevaratnam05}
G.~Seevaratnam, L.~Walker, E.~Ram\'{e} and S.~Garoff, \emph{J. Colloid Interf.
  Sci.}, 2005, \textbf{284}, 265--270\relax
\mciteBstWouldAddEndPuncttrue
\mciteSetBstMidEndSepPunct{\mcitedefaultmidpunct}
{\mcitedefaultendpunct}{\mcitedefaultseppunct}\relax
\EndOfBibitem
\bibitem[Sun \emph{et~al.}(2005)Sun, Feng, Gao, and Jiang]{sun05}
T.~Sun, L.~Feng, X.~Gao and L.~Jiang, \emph{Accounts Chem. Res.}, 2005,
  \textbf{38}, 644--652\relax
\mciteBstWouldAddEndPuncttrue
\mciteSetBstMidEndSepPunct{\mcitedefaultmidpunct}
{\mcitedefaultendpunct}{\mcitedefaultseppunct}\relax
\EndOfBibitem
\bibitem[Wei \emph{et~al.}(2007)Wei, Seevaratnam, Garoff, Ram\'{e}, and
  Walker]{WeiGaroff07}
Y.~Wei, G.~Seevaratnam, S.~Garoff, E.~Ram\'{e} and L.~Walker, \emph{J. Colloid
  Interf. Sci.}, 2007, \textbf{313}, 274--280\relax
\mciteBstWouldAddEndPuncttrue
\mciteSetBstMidEndSepPunct{\mcitedefaultmidpunct}
{\mcitedefaultendpunct}{\mcitedefaultseppunct}\relax
\EndOfBibitem
\bibitem[Wei \emph{et~al.}(2009)Wei, Ram\'{e}, Walker, and Garoff]{WeiGaroff09}
Y.~Wei, E.~Ram\'{e}, L.~Walker and S.~Garoff, \emph{J. Phys. Cond. Matt.},
  2009, \textbf{21}, 464126\relax
\mciteBstWouldAddEndPuncttrue
\mciteSetBstMidEndSepPunct{\mcitedefaultmidpunct}
{\mcitedefaultendpunct}{\mcitedefaultseppunct}\relax
\EndOfBibitem
\bibitem[Yue and Feng(2012)]{Yueetal12}
P.~Yue and J.~J. Feng, \emph{J. Non-Newton. Fluid.}, 2012, \textbf{189},
  8--13\relax
\mciteBstWouldAddEndPuncttrue
\mciteSetBstMidEndSepPunct{\mcitedefaultmidpunct}
{\mcitedefaultendpunct}{\mcitedefaultseppunct}\relax
\EndOfBibitem
\bibitem[Wang \emph{et~al.}(2015)Wang, Do-Quang, and Amberg]{Amberg}
Y.~Wang, M.~Do-Quang and G.~Amberg, \emph{Phys. Rev. E}, 2015, \textbf{92},
  043002\relax
\mciteBstWouldAddEndPuncttrue
\mciteSetBstMidEndSepPunct{\mcitedefaultmidpunct}
{\mcitedefaultendpunct}{\mcitedefaultseppunct}\relax
\EndOfBibitem
\bibitem[Usta \emph{et~al.}(2006)Usta, Butler, and Ladd]{Ladd05}
O.~B. Usta, J.~E. Butler and A.~C.~J. Ladd, \emph{Phys. Fluids.}, 2006,
  \textbf{18}, 031703\relax
\mciteBstWouldAddEndPuncttrue
\mciteSetBstMidEndSepPunct{\mcitedefaultmidpunct}
{\mcitedefaultendpunct}{\mcitedefaultseppunct}\relax
\EndOfBibitem
\bibitem[Graham(2011)]{GraamReview}
M.~D. Graham, \emph{Annu. Rev. Fluid Mech.}, 2011, \textbf{43}, 273--298\relax
\mciteBstWouldAddEndPuncttrue
\mciteSetBstMidEndSepPunct{\mcitedefaultmidpunct}
{\mcitedefaultendpunct}{\mcitedefaultseppunct}\relax
\EndOfBibitem
\bibitem[Ma and Graham(2005)]{MaGraam05}
H.~Ma and M.~D. Graham, \emph{Phys. Fluids}, 2005, \textbf{17}, 083103\relax
\mciteBstWouldAddEndPuncttrue
\mciteSetBstMidEndSepPunct{\mcitedefaultmidpunct}
{\mcitedefaultendpunct}{\mcitedefaultseppunct}\relax
\EndOfBibitem
\bibitem[Snoeijer and Andreotti(2013)]{snoeijer_moving_2013}
J.~H. Snoeijer and B.~Andreotti, \emph{Annu. Rev. Fluid Mech.}, 2013,
  \textbf{45}, 269--292\relax
\mciteBstWouldAddEndPuncttrue
\mciteSetBstMidEndSepPunct{\mcitedefaultmidpunct}
{\mcitedefaultendpunct}{\mcitedefaultseppunct}\relax
\EndOfBibitem
\end{mcitethebibliography}
\end{document}